\documentstyle[12pt]{article}
\newif\iffinalplot\finalplottrue
\newif\ifpictex\pictextrue
\renewcommand{\i}{\mbox{i}\/}

\catcode`\@=11
\def\undefine#1{\let#1\undefined}
\def\newsymbol#1#2#3#4#5{\let\next@\relax
 \ifnum#2=\@ne\let\next@\msafam@\else
 \ifnum#2=\tw@\let\next@\msbfam@\fi\fi
 \mathchardef#1="#3\next@#4#5}
\def\mathhexbox@#1#2#3{\relax
 \ifmmode\mathpalette{}{\m@th\mathchar"#1#2#3}%
 \else\leavevmode\hbox{$\m@th\mathchar"#1#2#3$}\fi}
\def\hexnumber@#1{\ifcase#1 0\or 1\or 2\or 3\or 4\or 5\or 6\or 7\or 8\or
 9\or A\or B\or C\or D\or E\or F\fi}

\font\tenmsa=msam10
\font\sevenmsa=msam7
\font\fivemsa=msam5
\newfam\msafam
\textfont\msafam=\tenmsa
\scriptfont\msafam=\sevenmsa
\scriptscriptfont\msafam=\fivemsa
\edef\msafam@{\hexnumber@\msafam}
\mathchardef\dabar@"0\msafam@39
\def\dashrightarrow{\mathrel{\dabar@\dabar@\mathchar"0\msafam@4B}}
\def\dashleftarrow{\mathrel{\mathchar"0\msafam@4C\dabar@\dabar@}}

\def\ulcorner{\delimiter"4\msafam@70\msafam@70 }
\def\urcorner{\delimiter"5\msafam@71\msafam@71 }
\def\llcorner{\delimiter"4\msafam@78\msafam@78 }
\def\lrcorner{\delimiter"5\msafam@79\msafam@79 }
\def\yen{{\mathhexbox@\msafam@55 }}
\def\checkmark{{\mathhexbox@\msafam@58 }}
\def\circledR{{\mathhexbox@\msafam@72 }}
\def\maltese{{\mathhexbox@\msafam@7A }}

\font\tenmsb=msbm10
\font\sevenmsb=msbm7
\font\fivemsb=msbm5
\newfam\msbfam
\textfont\msbfam=\tenmsb
\scriptfont\msbfam=\sevenmsb
\scriptscriptfont\msbfam=\fivemsb
\edef\msbfam@{\hexnumber@\msbfam}
\def\Bbb#1{{\fam\msbfam\relax#1}}
\def\widehat#1{\setbox\z@\hbox{$\m@th#1$}%
 \ifdim\wd\z@>\tw@ em\mathaccent"0\msbfam@5B{#1}%
 \else\mathaccent"0362{#1}\fi}
\def\widetilde#1{\setbox\z@\hbox{$\m@th#1$}%
 \ifdim\wd\z@>\tw@ em\mathaccent"0\msbfam@5D{#1}%
 \else\mathaccent"0365{#1}\fi}
\font\teneufm=eufm10
\font\seveneufm=eufm7
\font\fiveeufm=eufm5
\newfam\eufmfam
\textfont\eufmfam=\teneufm
\scriptfont\eufmfam=\seveneufm
\scriptscriptfont\eufmfam=\fiveeufm

\begin{document}
\title{
Completeness of ``Good'' Bethe Ansatz Solutions of a
Quantum Group Invariant Heisenberg Model
}
\author{G.~J\"uttner
        \thanks{Supported by DFG Sfb 288
                ``Differentialgeometrie und Quantenphysik''}
        \thanks{e-mail: juettner@omega.physik.fu-berlin.de}\\
        M.~Karowski
        \thanks{e-mail: karowski@omega.physik.fu-berlin.de}\\
        \parbox{10cm}
         {\begin{center}{\em
          Institut f\"ur  Theoretische Physik, FB~Physik \\
          Freie Universit\"at Berlin \\
          Arnimallee~14, 14195~Berlin \\
          Germany}
          \end{center}}
       }
\date{June 8, 1994}
%%\begin{document}
\maketitle
% short title: Completeness of ``Good'' Bethe Ansatz Solutions
% PACS: 75.10.JM,~~75.40~Fa
%
\begin{abstract}
  The $sl_q(2)$-quantum group invariant  spin 1/2 XXZ-Heisenberg model
  with open boundary conditions is  investigated by means of the Bethe
  ansatz.  As is well known, quantum groups for $q$ equal to a root of
  unity possess  a  finite  number of ``good''    representations with
  non-zero q-dimension and   ``bad'' ones with vanishing  q-dimension.
  Correspondingly, the state  space of  an invariant Heisenberg  chain
  decomposes into ``good'' and ``bad'' states. A ``good'' state may be
  described by a  path of only ``good''  representations.  It is shown
  that the  ``good''  states are given  by  all ``good''  Bethe ansatz
  solutions with roots restricted to the first periodicity strip, i.e.
  only  positive  parity strings (in   the  language of Takahashi) are
  allowed. Applying  Bethe's string counting technique completeness of
  the ``good'' Bethe states is proven, i.e.  the same number of states
  is  found  as  the     number  of  all  restricted  path's   on  the
  $sl_q(2)$-Bratteli diagram.  It    is    the first  time that      a
  ``completeness" proof for an  anisotropic quantum invariant  reduced
  Heisenberg model is performed.
\end{abstract}
\section{Introduction}\label{intro}
The Bethe  ansatz  method  has been   applied  to a  large  number  of
integrable  models  as   one-dimensional  Heisenberg spin  chains  and
statistical lattice models  in two   dimensions. The underlying   Yang
Baxter Algebra is responsible  for the integrability of these  models.
Yang Baxter Algebras are  related to new mathematical structures often
referred to as quantum groups which were introduced by Drinfeld (1986)
and  Jimbo (1985).   On the  other  hand,  these  quantum  groups have
attracted attention as a   powerful  tool for studying   properties of
solvable systems.

The isotropic XXX-Heisenberg model solved  by Bethe (1931) corresponds
to  a rational solution of the  Yang Baxter equation (Baxter 1982) and
is $SU(2)$  symmetric.  A  deformation of  the  XXX model leads  to an
anisotropic XXZ-Heisenberg model related  to trigonometric Yang Baxter
solutions.  A  version of the  model with  open boundary conditions is
quantum group invariant.   The Bethe ansatz  method was used to  solve
the quantum   invariant  spin chain for   open boundary  conditions by
several  authors    (see e.g.  Alcaraz et   al   1987, Cherednik 1984,
Sklyanin 1988, Mezincescu and  Nepomechie 1991, Martin and  Rittenberg
1992, Destri and de Vega 1992, Foerster and Karowski 1993).  A quantum
group  invariant version  with  periodic boundary conditions has  been
constructed and analyzed  in (Karowski and  Zapletal  1993, 1994).  We
restrict here our interest to a chain with open boundary conditions.

For generic  values of $q$ the representations  of $sl_q(2)$ are known
to be equivalent to the ordinary $SU(2)$ representations (Luztig 1989,
Rosso 1988).  However,    this   correspondence holds  only   if   the
deformation parameter   $q$  is not  a  root   of unity
$(q^r\neq\pm 1, r=\mbox{integer})$.

In  contrast  to this generic case   the  quantum group representation
theory  for $q^r=\pm 1$ is  more complicated  (Luztig 1989, Reshetikin
and Smirnov 1989, Pasquier  and  Saleur  1990, Reshetikin  and  Turaev
1991),  because  there   exits  two  types  of  representations.   The
representations   with    non-zero     q-dimension    (see     section
\ref{sec:tensor-basis}     equation(\ref{q-dimension}))   are   called
``good'' (Reshetikin  and  Smirnov 1989) or  of  type-II (Pasquier and
Saleur 1990).  Moreover, all of them have positive q-dimension, if and
only if $q=e^{i\pi/r}$.  They  are irreducible  and  possess the  same
structure as the usual  $SU(2)$ ones.  There  are only a finite number
of     ``good''   representations,    namely,     those    with   spin
\mbox{$j=0,1/2,1,\dots,r/2-1$}   if \mbox{$q^r=\pm  1$}.   Those  with
vanishing q-dimension  called ``bad'' or  type-I representations. Some
of them are irreducible and have  spin $j=(nr-1)/2$. The others can be
described  as  a mixing of  two  representations of  the generic case.
They   are  reducible but   indecomposable.    This phenomenon   has a
consequence   for  the  eigenstates  of   an quantum  group  invariant
Hamiltonian.   If $q$  approaches  a root  of unity,  some eigenstates
which correspond to ``bad'' representations  become dependent and will
mix. Due to the coincidence  of two originally independent eigenstates
the whole eigenspace  is {\em not}  complete, i.e. the Hamiltonian may
not be completely diagonalizable. This  phenomenon was investigated by
Bo-Yu Hou et al (1991).

The existence   of  ``good'' and ``bad''  representations  implies the
decomposition of the state space of an Heisenberg chain with $N$ sites
into  ``good'' and ``bad''  states. We consider  the state iteratively
fused  by the  $N$  spin  1/2-representations. The  ``good''  ones are
characterized  by the condition  that all intermediate representations
have also  to be ``good'' ones.  This  means  that the ``good'' states
are  described by   a    restricted  ($j\le   r/2-1$)  path    on   an
$sl_q(2)$-Bratteli diagram.  A projection of the full Hilbert space to
the subspace of  ``good'' states is analog to  the  restriction of the
solid-on-solid model (SOS) to the so-called RSOS  model (Andrews et al
1984), where   the local  height   variables of   the  SOS model   are
restricted  to  the finite set  $(1,2,\dots   ,r-1)$.  This provides a
connection  to the minimal models of  conformal field theories related
to critical  phenomena    of   2D systems  with   second  order  phase
transitions.  Indeed, the   analysis of  finite  size  corrections  of
quantum   group  invariant XXZ  spin  chains  (Hamer  et  al 1987) and
RSOS-models (Karowski  1988) lead  to  conformal charges  smaller than
one.

In section \ref{sec:tensor-basis} we  present  the model in  terms  of
Pauli matrices and Temperly-Lieb  operators. In addition we write some
$sl_q(2)$-formulas  which   will   be    used  later.     In   section
\ref{sec:path-basis}  we formulate the  model   in terms of  the  path
basis. This formulation is closely related to the RSOS model. It leads
to an explicit ``quantum group reduction''.

In  section \ref{sec:bae-string} we  investigate the model by means of
the Bethe  ansatz method.   The  eigenstates  and eigenvalues  of  the
Hamiltonian are  described  by sets  of  parameters, the  Bethe ansatz
roots. Such set of roots satisfy a system  of algebraic equations, the
Bethe ansatz  equations (BAE).  As a  generalization of Bethe's (1931)
work  Takahashi (1971, 1972) introduced   for the XXZ-Heisenberg model
the general  string picture.  This means   that a solution of  the BAE
consists of a series of strings in the form
$$
\lambda =\Lambda  + i m,
\quad\Lambda =\mbox{real},\quad
m=-M,-M+1,\dots ,M,
\quad M=0,\pm 1/2, \pm 1, \pm 3/2, \dots\quad
$$
with a positive parity or strings with a negative parity
$$
\lambda =\Lambda +i r/2+i m.
\quad\Lambda =\mbox{real},\quad
m=-M,-M+1,\dots ,M,
\quad M=0,\pm 1/2, \pm 1, \pm 3/2, \dots\quad
$$
In this   context  the question  arises  how  ``good'' and  ``bad''
eigenstates are  determined   by  the solutions of   the  Bethe ansatz
equations.  We conjecture that {\em  ``good'' states are exactly given
  by strings of  positive  parity restricted to the  first periodicity
  strip $\Im\lambda <r/2$ and whose total spin fulfill $j\leq r/2-1$.}
We have no rigorous proof for this conjecture, but we are able to show
the following coincidence.  The   total number of all possible  string
configurations of restricted positive parity coincides with the number
of ``good'' states in the path picture of  the model, which is counted
by the number of all  ``good'' path's, i.e.   all restricted path's on
the spin $1/2$ $sl_q(2)$-Bratteli  diagram.  This completeness will be
shown in section \ref{sec:bae-string}.  Such ``proofs of completeness"
of Bethe  ansatz  solutions  are   well known for  models  with  group
symmetry.  Already Bethe applied  this procedure to the XXX-Heisenberg
model.   For  other  models  see E\ss  ler  et  al   (1992a)  for  the
$SU(2)\times SU(2)$ symmetric Hubbard model  and Foerster and Karowski
(1992, 1993)  for  the $spl(2,1)$-t-J-model.    After having completed
this paper we received a preprint (Kirillov and Liskova 1994) in which
a ``completeness proof'' is treated for  XXZ spin chains with periodic
boundary conditions which have no quantum group symmetry.  In contrast
to these models,  we here present for the  first time a ``completeness
proof'' for an anisotropic  quantum group symmetric (spin $1/2$) model
where the number of ``good'' states is smaller than $2^N$.

\section{Tensor basis description of $H$}
\label{sec:tensor-basis}

The $sl_q(2)$ invariant XXZ  Hamiltonian with open boundary  condition
can be expressed in terms of Pauli matrices
\begin{equation}
\label{xxz-hamilton}
H= \sum_{i=1}^{N-1}\left( \sigma _i^x \sigma _{i+1}^x +
  \sigma _i^y \sigma _{i+1}^y +
  \frac{q+q^{-1}}{2}\left(\sigma _i^z \sigma _{i+1}^z -1 \right) \right)
  +\frac{q-q^{-1}}{2}\left( \sigma_1^z-\sigma_N^z \right).
\end{equation}
This expression may be rewritten in terms of Temperly Lieb operators
%(see e.g. Pasquier and Saleur 1990, Meljanac et al 1991)
\begin{equation}\label{TL}
T=
% \sigma_i^x \sigma_{i+1}^x + \sigma _i^y \sigma _{i+1}^y +
%  \frac{q+q^{-1}}{2} (\sigma _i^z \sigma _{i+1}^z -1 )
%  +\frac{q-q^{-1}}{2}( \sigma_i^z-\sigma_{i+1}^z )=
\left(\matrix{0&0&0&0\cr 0&q^{-1}&-1&0\cr 0&-1&q&0\cr 0&0&0&0}\right)
\end{equation}
which satisfy the relations
\begin{eqnarray}
T_k T_{k+1} T_k &=& T_k, \nonumber \\
T_k^2 &=& (q+q^{-1}) T_k, %\quad \beta =q+q^{-1}
  \label{temperly-lieb-relation} \\
T_k T_l &=& T_l T_k, \quad \mid k-l\mid\geq 2. \nonumber
\end{eqnarray}
The Hamiltonian $H$ reads
\begin{equation}
  H = -2 \sum_{k=1}^{N-1} T_k.
\end{equation}
This formula suggests the notation of open boundary conditions.

The   model is quantum group  symmetric,    since the Hamiltonian  $H$
commutes with the  generators $S^\pm , S^z$  of the q-deformed algebra
$U_q(sl(2))$
\begin{equation}
  [H,S^\pm ]=0, \quad [H,S^z]=0.
\label{h-s-commutator}
\end{equation}
We list here some formulas  which will be  used later.  The generators
possess the following properties
\begin{equation}
[S^+, S^-]=[2 S^z]_q, \quad q^{S^z}S^\pm q^{-S^z}=q^{\pm 1} S^\pm .
  \label{s-commutator}
\end{equation}
The q-number $[x]_q$ is defined as
\begin{equation}
[x]_q=\frac{q^{x}-q^{-x}}{q-q^{-1}}.
  \label{deformation-symbol}
\end{equation}
In the limit $q\to  1$  one finds $[x]_q=x$ and   (\ref{s-commutator})
tends to the  usual relations of $SU(2)$.  For  $q$ equal to a root of
unity one has in addition
\begin{equation}
(S^\pm )^r=0\quad{\rm for}\quad q^r=\pm 1.
\end{equation}
The underlying Hopf algebra  structure (Drinfeld 1986, Jimbo 1985)  is
described by coproduct, antipode and counit
\begin{equation}
\begin{array}{rclrcl}
  \Delta (q^{\pm S^z}) &=& q^{\pm S^z}\otimes q^{\pm S^z},\quad &
  \Delta (q^{S^\pm}) &=& q^{S^z}\otimes S^\pm + S^\pm\otimes q^{-S^z} \\
  {\gamma } (q^{\pm S^z}) &=& q^{\mp S^z}, &
  {\gamma } (q^{S^\pm}) &=& -q^{\pm 1} S^\pm, \\
  \varepsilon (q^{\pm S^z}) &=& 1, &
  \varepsilon (q^{S^\pm}) &=& 0.
\end{array}
\label{hopf-structure}
\end{equation}

The configuration space
$({\Bbb C}^2)^{\otimes N}$
of a quantum spin $1/2$ chain
with $N$ lattice points has the natural tensor basis
$$
|\underline\alpha\rangle =
|\alpha_N,\dots,\alpha_1\rangle =
|\alpha_N\rangle\otimes\cdots\otimes|\alpha_1\rangle
$$
with $|-\frac 12\rangle=(\downarrow )={0\choose 1}$ (spin down)
and $|\frac 12\rangle=(\uparrow )={1\choose 0}$ (spin up).
The representation of the quantum group generators
on the configuration space
%$({\Bbb C}^2)^N$
is given by
\begin{eqnarray}
q^{S^z} &=& q^{\sigma^z/2}\otimes\cdots\otimes q^{\sigma^z/2}, \\
S^\pm   &=& \sum_i^N S^\pm_i, \\
S^\pm_i &=& q^{\sigma^z/2}\otimes\cdots\otimes q^{\sigma^z/2}
            \otimes\sigma^\pm_i/2\otimes
            q^{-\sigma^z/2}\otimes\cdots\otimes q^{-\sigma^z/2}.
\label{s-generators}
\end{eqnarray}

The q-deformed Casimir operator is
\begin{equation}\label{casimir}
S^2 =S^-S^+ +[S^z+1/2]_q ^2 -[1/2]_q ^2.
\end{equation}
For generic values of $q$ all representations $\rho_j$ are irreducible
and highest weight ones with spin  $j=0,1/2,1,\dots$ (see e.g. Lusztig
1989, Rosso 1988).  If $q$ tends to a root  of unity ($q^r=\pm 1$) the
Casimir eigenvalue
\begin{equation}
S^2=[j+1/2]^2_q-[1/2]^2_q
  \label{casimir-highest-weight}
\end{equation}
can take identical values for different spins $j$ and $j'$ in case
\begin{equation}
  j'=j+nr \quad\mbox{or}\quad  j'=r-1-j+nr, \quad n\in {\Bbb  Z}.
  \label{mixed-j-j}
\end{equation}
The representations $\rho_j$ and $\rho_{j'}$,  which are different for
generic  $q$,  mix for       $q^r=\pm 1$  and build    reducible   but
indecomposable representations.      Together    with the  irreducible
representations with spin  $j=(nr-1)/2$   they  are characterized   by
vanishing q-dimension defined by
\begin{equation}
\mbox{dim}_q\,\rho ={\rm tr}_{V^\rho}(q^{-2S^z})
  \label{q-dimension}
\end{equation}
where $V^{\rho}$  is   the representation  space.  Therefore they  are
called ``bad'' or  type  I representation.  On  the  other hand, there
exits   a  finite number  of  representations  which are in one-to-one
correspondence to $SU(2)$ representations.  They are irreducible, have
spin $j=0,1/2,\dots,r/2-1$ and non-vanishing q-dimension
\begin{equation}
d_j=\mbox{dim}_q\rho_j =[2j+1]_q.
\end{equation}
They are   called   ``good''   or   type-II representations.     Their
q-dimension is always positive if $q=e^{i\pi/r}$.

Reshetikhin and Turaev (1991) have shown that for the ``good'' and the
``bad'' representation spaces the following structure holds
\begin{eqnarray}
V_{good}\otimes V_{good}&=&\left(\bigoplus V_{good}\right)
\oplus\left(\bigoplus V_{bad}\right) ,\\
V_{good}\otimes V_{bad}&=&\bigoplus V_{bad}.
\end{eqnarray}
This means that  ``unitarity'' in  the   ``good'' subspace (see   also
Reshetikin and Smirnov   1989)  is fulfilled for   ``good''  covariant
operators
\begin{equation}
\langle\ good\,'\mid A_{j_{good}}B_{k_{good}}\mid good\ \rangle=
\sum_{good\,''}
\langle\ good\,'\mid A_{j_{good}}\mid good\,''\ \rangle
\langle\ good\,''\mid B_{k_{good}}\mid good\ \rangle.
\end{equation}

In the next  section the ``good''  representations in the  state space
$({\Bbb  C}^2)^{\otimes  N}$    of   the XXZ-Heisenberg  model     are
characterized in  terms of the ``path  picture''.  For $q$  equal to a
root  of unity the state space  may be reduced to  the subspace of all
``good'' states. Changing the metric  in this subspace the Hamiltonian
(\ref{xxz-hamilton}) will become selfadjoint.

\section{Path basis formulation}
\label{sec:path-basis}

For  generic   values   of   the   deformation   parameter  $q$    the
representations of $sl_q(2)$ are  irreducible and classified  as those
of the  undeformed  group $SU(2)$. The   space $V^j$  of the spin  $j$
representation  is  spanned  by     a  set of basis   vectors
$\mid j,m\ \rangle$, $m=j,j-1,\dots,-j$ with
\begin{eqnarray}
S^\pm \mid j,m\ \rangle&=&
\sqrt{[j\mp m]_q [j\pm m+1]_q} \mid j,m\pm 1\ \rangle
\label{jm-basis}\\
S^z \mid j,m\ \rangle&=& m \mid j,m\ \rangle. \nonumber
\end{eqnarray}
The tensor product  space  $V^{j_1}\otimes V^{j_2}$ decomposes  into a
direct sum of irreducible spaces
$$
V^{j_1}\otimes V^{j_2} =
\bigoplus_{j=\mid j_1-j_2\mid}^{j_1+j_2} V^j
$$
given  by  Clebsch-Gordan  coefficients   (see  e.g.  Kirillov  and
Reshetikin 1989)
\begin{equation}
\mid j,m\ \rangle_{j_1,j_2}=\sum_{m_1,m_2}
 \mid j_1m_1\ \rangle\otimes \mid j_2m_2\ \rangle
 \left|\begin{array}{ccc} j& j_2& j_1 \\ m& m_2& m_1 \end{array}\right|_q.
\end{equation}
By  successive fusion of the $N$ spin $1/2$ states associated to the
lattice sites we construct the state
\begin{equation}\label{trans}
|\ \underline j,m\ \rangle =\sum_{\underline\alpha}|\ \underline\alpha\
\rangle\langle\ \underline\alpha\ |\ \underline j,m\ \rangle
\end{equation}
which is labeled by the ``path'' of spins $\underline j =
(j_N, j_{N-1}, \dots, j_2, j_1=1/2)$ and the magnetic quantum number
$m=\sum_i\alpha_i$. The matrix element is a product of Clebsch-Gordan
coefficients
\begin{equation}\label{trans-matrix}
\langle\ \underline\alpha\ |\ \underline j,m\ \rangle =
\sum_{m_2,\dots,m_{N-1}}
 \left|\matrix{ j_N& 1/2& j_{N-1} \cr m&\alpha_N & m_{N-1}} \right|_q
\cdots
 \left|\matrix{ j_3& 1/2& j_2 \cr m_3&\alpha_3 & m_2} \right|_q
 \left|\matrix{ j_2& 1/2& 1/2 \cr m_2&\alpha_2 & \alpha_1 }\right|_q
\end{equation}
The spins $j_k$ are restricted by the fusion rule
\begin{equation}
j_{k+1}=j_k\pm 1/2
  \label{fusion-rule}
\end{equation}
and $j=j_N$ describes the total spin of the state.
The space $V_{1/2}^{\otimes N}$ is decomposed into
a direct product of a space described by paths $W_j$ and a space $V_j$
where the generators $S^\pm, S^z$ act
\begin{equation}
  V^{\otimes N}=\sum_j W_j\otimes V_j.
\end{equation}
Quantum group invariant operators as the Hamiltonian
only act in the path space $W_j$. By the Wigner-Eckart theorem
the magnetic quantum number $m$ is not be changed and in addition the
path space (or reduced) matrix elements do not depend on $m$.
Therefore we will omit it in the following.

For example the Temperly Lieb operators (\ref{TL}) in path space act as
\begin{equation}\label{temperly-lieb-action}\\
T_k\mid\dots,j_k,\dots\rangle=
\delta_{j_{k-1}j_{k+1}}\sum_{j'_k=j_{k+1}\pm 1/2}
\mid \dots,j'_k,\dots\rangle \frac{\sqrt{d_{j_k}d_{j'_k}}}{d_{j_{k+1}}}.
\end{equation}
Thus, we easily obtain the matrix in path space of the Hamiltonian $H$,
which decomposes into different blocks for different total spin $j$.

Now we turn to the case where the deformation parameter $q$
is a root of unity
(\mbox{$q^r=\pm 1$}). In the path picture it is very simple to
characterize the
``good'' states. From section \ref{sec:tensor-basis} it is obvious
that the ``good'' states are
given by all restricted paths
\begin{equation}\label{good-path}
|\ \underline j_{good}\ \rangle=\mid j_{n},
\dots,j_{1}\ \rangle,\quad
2j_k+1<r,\quad k=1,\dots,N.
\end{equation}

For generic values of $q$ the number of all states with total spin $j$ is
equal to
the number of all possible unrestricted paths (\ref{fusion-rule})
on the $sl(2)$-Bratteli diagram
\begin{equation}
\Gamma_j={N\choose N/2-j}-{N\choose N/2+1+j}.
  \label{all-path-number}
\end{equation}
For $q^r=\pm 1$ the number of all ``good'' states with total spin
$j<r/2-1$ is equal to
the number of all possible restricted paths (\ref{good-path})
on the cut $sl_q(2)$-Bratteli diagram
\begin{equation}
\Omega_j=\sum_{k=-\infty}^{\infty}\Gamma_{j+rk}.
  \label{good-path-number}
\end{equation}
In the next section, it turns out that this number coincides with the
number of, what we will introduce, the ``good'' Bethe ansatz states.
Note that only a finite number of terms contribute, because
the binomial coefficient ${m\choose n}$ is defined to
be zero for integer $n<0$ or $n>m$.

The Hamiltonian does not lead to
a transition from a ``good'' state to a ``bad'' one.
This follows from relation (\ref{temperly-lieb-action}). The only
possible ``good'' path's $\mid\underline j_{good}\ \rangle$ which lead
by an action
of $H=-2\sum T_k$ to a ``bad'' state $\mid\underline j_{bad}'\ \rangle$ by
\begin{equation}
T_k\mid\underline j_{good}\ \rangle=c\mid\underline j_{good}\ \rangle+
c'\mid\underline j_{bad}'\ \rangle
\end{equation}
must have one or more values $j_{k+1}=j_{k-1}=j_k+1/2$ with
$2j_{k+1}+1=r-1.$
Then the ``bad'' path would have the value of $j'_k=j_k+1=(r-1)/2$.
But the coefficient $c'$ vanishes because of eq.~(\ref{temperly-lieb-action})
and $d_{j'_k}=[r]_q=0$. That means, the matrix $H$ in path space
decomposes into a ``good'' submatrix and a ``bad'' part.
Note that these arguments cannot be directly translated to the tensor
picture, because the transition matrix elements (\ref{trans-matrix})
become singular for ``bad'' states.

We now introduce a new scalar product in the ``good'' subspace in the path
formulation, such that the Hamiltonian becomes selfadjoint
\begin{equation}
\langle\ \underline j',m'\mid\underline j,m\ \rangle
=\delta_{m'm}\delta_{\underline j'\underline j}.
\end{equation}
Note that this scalar product coincides with the natural one in the
tensor picture only for real values of $q$.
The selfadjointness of the Hamiltonian follows from
eq.~(\ref{temperly-lieb-action})
\begin{equation}
H^\dagger=H\quad{\rm for}\quad q=e^{i\pi/r},\ r=2,3,\dots
\end{equation}
since the q-dimensions $d_j=[2j+1]_q$ are positive for $2j+1<r$.

\section{Bethe ansatz method}
\label{sec:bae-string}

The     eigenstates  and    eigenvalues   of    the    XXZ-Hamiltonian
(\ref{xxz-hamilton})  with open boundary  conditions are  described by
sets of distinct spectral parameters $\{\lambda_1,\dots ,\lambda_l\}$,
the roots of the Bethe ansatz equations (BAE)
\begin{equation}
\left(\frac{\sinh\gamma (\lambda_j+\i /2)}
           {\sinh\gamma (\lambda_j-\i /2)}
\right)^{2N} =
\prod_{k=1, k\neq j}^l
\frac{\sinh\gamma (\lambda_j-\lambda_k+\i )}
     {\sinh\gamma (\lambda_j-\lambda_k-\i )}
\frac{\sinh\gamma (\lambda_j+\lambda_k+\i )}
     {\sinh\gamma (\lambda_j+\lambda_k-\i )},
\label{bae}
\end{equation}
where $q=\exp(\i\gamma)$.  Because  of the open boundary conditions it
is   sufficient  to consider  only   roots with   positive real  parts
$0<\Re\lambda_k<\infty$.  The energy is
\begin{equation}
  E=-4\sum_{k=1}^l\frac{\sin^2\gamma}{\cosh 2\gamma\lambda_k
    -\cos\gamma}.
  \label{bae-energy}
\end{equation}
The total spin $j$ of an eigenstate is  related to the number of roots
$l$ by
\begin{equation}
j=N/2 -l,\quad l=0,\dots ,N/2
\end{equation}
where the lattice  length $N$ is  assumed to be  even.  Because of the
quantum group invariance the Bethe ansatz solutions are highest weight
states (Destri and de Vega 1992).  The  aim of our investigation is to
find  a  concrete  criterion  for the  Bethe  ansatz  solutions to  be
``good'' states in the sense of section 2 and 3.

A detailed  numerical  analysis  of  the  BAE  motivated also  by  the
observation of Destri  and de Vega  (1992) that when  $q$ approaches a
root of unity  the appearance of a  ``bad'' state correspond to a root
tending  to infinity (see also  Appendices  A and B)   leads us to the
following

\medskip\noindent{\bf\underbar{Conjecture 1}: } {\em For
  \mbox{$q=\exp(\i\pi /r)$} a Bethe ansatz state is the highest weight
  vector of a ``good'' representation (in the sense of section
  \ref{sec:tensor-basis} and \ref{sec:path-basis} see
  eq.~(\ref{good-path}), if and only if
\begin{itemize}
\item[(i)]
  the total spin $j$ is restricted by $2j+1<r$, i.e.
  the number of roots $l$ must be larger than $(N+1-r)/2$,
\item[(ii)]
  the roots are restricted to the first periodicity
  strip $\mid \Im\lambda_k\mid < r/2$.
\end{itemize}
}
The  first  condition (i) is obvious   from the definition of ``good''
states.  We have no rigorous proof for  the second condition (ii), but
we can  show  completeness in the  sense  that the  number of ``good''
Bethe ansatz solutions defined by   Conjecture 1 leads to the  correct
number of all ``good'' states  on the lattice  counted by all paths on
the Bratteli diagram  (see eq.~(\ref{good-path-number})).  To  "prove"
{\bf Conjecture 1}  we proceed  as  follows.   We classify  the  Bethe
ansatz  roots by means of  Takahashi's string picture. A configuration
of strings is given  by sets  of integers.  In  {\bf Conjecture  2} we
give  upper bounds for these  integers.   This is  also  for the group
symmetric case a nontrivial  (but simpler) problem.  The upper  bounds
are smaller than a naive estimate  would suspect.  The reason for this
phenomenon  is   that the   string  picture   is  only a    very rough
approximation. The exact set of roots are given by deformed (sometimes
even degenerated) strings. This leads to the  fact that less roots are
possible than  the  exact string picture would  allow.   For the group
symmetric  case   already Bethe (1931)  was   able solve this problem,
because there is a natural way  to fix the  bounds in order to get the
correct number    of states. This is  much   more complicated  for the
quantum  group case.  Only guided by  a lot of numerical calculations,
we were able to solve this problem.

It is convenient   to  classify  solutions  by the    so-called string
hypothesis (Takahashi 1971, 1972).   Any solution of the BAE  consists
approximatively of a series of strings in the form
\begin{equation}
\lambda_M =\Lambda_M + \i m, \quad
m=-M,-M+1,\dots ,M,
\quad M=0,\pm 1/2, \pm 1, \pm 3/2, \dots
  \label{string-positive}
\end{equation}
with    positive  parity     or     strings with     negative   parity
$\lambda=\Lambda+\i r/2+\i m$.  According to Conjecture 1 (ii) we must
take into account only the positive type (\ref{string-positive}).

A Bethe vector is characterized by the sets of real string centers
\begin{equation}
\{\{\Lambda_0\},\{\Lambda_1\},\dots \}\quad{\rm where}\quad
\{\Lambda_M\} = \{\Lambda_{M,1},\dots ,\Lambda_{M,\nu_M}\}.
\end{equation}
The  length of   a string is   $2M+1$  and $\nu_M$  is  the number  of
$(2M+1)$-strings  with  different centers $\Lambda_{M,k}$.   The total
number of roots writes as
\begin{equation}
l = \sum_{M=0} (2M+1)\nu_M.
  \label{root-number}
\end{equation}
As usual we rewrite the Bethe ansatz equations in terms of the strings
centers
\begin{eqnarray}
V^{2N}_{M+1/2}(\Lambda_{M,i}) &=&
\prod_{m=1}^{2M} V^2_m(2\Lambda_{M,i})\nonumber\\
&& \times
\prod_{M'}\prod_{k=1 \atop \{M,i\}\neq\{M',k\}}^{\nu_{M'}}
V_{M,M'}(\Lambda_{M,i}-\Lambda_{M',k})
\ V_{M,M'}(\Lambda_{M,i}+\Lambda_{M',k}),
  \label{bae-string-product}
\end{eqnarray}
where
\begin{equation}
V_{M,M'}(\lambda ) =\prod_{m=\mid M-M'\mid}^{M+M'}
V_{m}(\lambda )V_{m+1}(\lambda )\quad{\rm and}\quad
V_m(\lambda )=
\frac{\sinh\gamma (\lambda +\i m)}{\sinh\gamma (\lambda -\i m)}.
\end{equation}
Taking the logarithm of eq.~(\ref{bae-string-product}) sets of
integers $\{Q_{M}\}$ occur which determine the solutions
\begin{eqnarray}
\lefteqn{
2N\Psi_{M+1/2}(\Lambda_{M,i}) = 2\pi Q_{M,i} +
\sum_{m=1}^{2M}2\Psi_m(2\Lambda_{M,i})}\nonumber\\
&&\qquad+\sum_{M'}\sum_{k=1 \atop \{M,i\}\neq\{M',k\}}^{\nu_{M'}}
\Big(
\Psi_{M,M'}(\Lambda_{M,i}-\Lambda_{M',k})+
\Psi_{M,M'}(\Lambda_{M,i}+\Lambda_{M',k})
\Big)
  \label{bae-string}
\end{eqnarray}
where
\begin{eqnarray}
\Psi_{M,M'}(\lambda ) &=& \sum_{m=\mid M-M'\mid}^{M+M'}
\big(\Psi_{m}(\lambda )+\Psi_{m+1}(\lambda) \big)\quad {\rm and}\\
\Psi_m(\lambda )&=& 2 \arctan
   \left(\cot (\gamma m)\tanh (\gamma\lambda )\right), \quad
m>0,\quad\Psi_0(\lambda )= 0 .
\end{eqnarray}
A solution of the BAE
is parameterized by a configuration of distinct integers
\begin{equation}
0<Q_{M,1}<Q_{M,2}<\dots <Q_{M,\nu_M}\leq Q_M^{\max},
\quad M=0,1/2,1,\dots
\label{string-values}
\end{equation}
which can  be used to  count the  total number  of possible solutions.
The upper bounds $Q_M^{\max}$ are defined  by the following condition.
All     roots     corresponding   to   the     integers     fulfilling
(\ref{string-values}) are ``good'' ones in the  sense of Conjecture 1.
However,  if one  $Q_{M,\nu_M}=Q_M^{\max}+1$   then one root would  be
pushed to  $\infty$ or  it would return   as a ``bad'' one  with $|\Im
\lambda|\ge  r/2$.  This is meant in  the sense of deforming the model
with respect   to $q$.  The counting    of the ``good"  states  may be
performed by the following

\medskip\noindent{\bf\underbar{Conjecture 2}: } {\em
For ``good" Bethe ansatz solutions:
\begin{itemize}
\item[(iii)]
 the upper bounds in relation (\ref{string-values}) are given by
 \begin{equation}\label{qmax}
 Q_M^{\max} = 2j+\nu_M+2\sum_{M'>M}2(M'-M)\nu_{M'}-G_M,\quad
 G_M = \max(2j+2M+3-r,0)
 \end{equation}
\item[(iv)]
 and the string length is restricted by
 \begin{equation}
  2M+1\le r-2.
  \label{string-max-length}
 \end{equation}
\end{itemize}
}
Note that  this condition (iv)    is  stronger than expected  from
Conjecture 1 (ii) and (\ref{qmax}), which  would allow also strings of
length $r-1$.  However, (iv) follows from (iii), since for the maximal
string the   sum   in eq.~(\ref{qmax})   is   empty and   by  relation
(\ref{string-values})
$$
\nu_{M_{\max}}\le Q^{\max}_{M_{\max}}=2j+\nu_{M_{\max}}-G_{M_{\max}}\le
\nu_{M_{\max}}-2M_{\max}-3+r.
$$
Note also that a naive estimate would lead  to a larger upper bound
than that  of   (iii), namely to   the  largest integer smaller   than
$Q_M^\infty$, which is obtained by putting $\Lambda_{M,i}\to\infty$ in
eq.~(\ref{bae-string}) (using $\Psi_m(\infty )=\pi -2\gamma m$)
\begin{equation}
Q_M^{\infty} = 2j+\nu_M+2\sum_{M'>M}2(M'-M)\nu_{M'}+
(2M+1)\left( 1-\frac{1}{r}(N-2l+2M+2)\right).
  \label{string-infty}
\end{equation}
The upper bound of (iii) is  obviously smaller than that obtained from
this equation,  even for the isotropic case  $q=1$ or $r=\infty$ where
$Q^{\max}_M(q=1)=Q^{\infty}_M-2M-1$, already used by Bethe (1931) with
\mbox{$J(M,M')=2\min(M,M')+1-\delta_{M,M'}1/2$} and
\begin{eqnarray}
Q^{\max}_M(q=1) &=& 2j+\nu_M+2\sum_{M'>M}2(M'-M)\nu_{M'} \\
&=& N-2\sum_{M'}\nu_{M'}J(M,M').
\end{eqnarray}
We are not able to prove  (iii) rigorously, however, we performed many
numerical calculations (see Appendices A and B)  and justify it by the
following counting of ``good" states leading to the correct result.

We  must  solve the  following  combinatorial problem  to compute  the
number of Bethe ansatz  states.  The integer $Q_M^{\max}$  denotes the
number of vacancies for a string of given length $2M+1$. The number of
possible configurations (\ref{string-values}) is ${Q_M^{\max}  \choose
  \nu_M}$  and  therefore the number of  combinations  of a  given set
$\{\nu_M\}$ reads
\begin{equation}
Z(N,\{\nu_M\},r)=\prod_{M}{Q_M^{\max} \choose \nu_M}.
\label{z-nu-prod}
\end{equation}
Now we obtain the total  number of Bethe  ansatz states with fixed $l$
by taking the sum over all configurations $\{\nu_M\}$
\begin{equation}
Z(N,l,r)=\sum_{\{\nu_M\}}Z(N,\{\nu_M\},r)\quad{\rm with\ fixed}\quad
l=\sum_M (2M+1)\nu_M.
\label{z-fixed-l-number}
\end{equation}
Extending the calculations made by Bethe (1931) one can show
for $M>0$
\begin{equation}
Q_M^{\max}(N,\{\nu_M\},r) = Q_{M-1/2}^{\max}(N-2\mu ,\{\nu_M'\},r-1),
\quad \mu = \sum_M \nu_M ,\quad \nu_M'=\nu_{M+1/2}.
\end{equation}
This implies with eqs.~(\ref{z-nu-prod}) and (\ref{qmax})
\begin{equation}
Z(N,\{\nu_M\},r) ={Q_0^{\max}\choose \nu_0} Z(N-2\mu,\{\nu_M'\},r-1)
\quad{\rm with}\quad
Q_0^{\max} = N-2\mu +\nu_0-G_0(r).
\label{z-nu-rec-part}
\end{equation}
We introduce  the partial number  of  configurations depending on  the
number of strings $\mu$
\begin{equation}
Z(N,l,\mu,r)=\sum_{\Sigma (2M+1)\nu_M=l \atop \Sigma\nu_M =\mu}
Z(N,\{\nu_M\},r).
\label{z-string-sum}
\end{equation}
{}From  eq.~(\ref{z-nu-rec-part}) we  obtain   the following  recurrence
relation
\begin{equation}
Z(N,l,\mu,r)=\sum_{\nu_0 =0}^{\mu -1}{Q_0^{\max}\choose \nu_0}
             Z(N-2\mu,l-\mu,\mu-\nu_0,r-1).
  \label{recurrence-relation}
\end{equation}
This relation differs from that   of Bethe for   the XXX-model by  the
additional     dependence    on   $r$.  Note    that    the   equation
(\ref{recurrence-relation}) holds for \mbox{$\mu <l$}.

The initial  values of this  recurrence relations are given by $l=\mu$
where only  real  roots exist  ($\nu_0=l$  and $\nu_M=0$  for  $M>0$),
therefore from eqs.~(\ref{z-nu-prod}) and (\ref{z-string-sum})
\begin{eqnarray}
Z(N,l,\mu=l,r)&=&
{Q^{\max}_0\choose l}={2j+l-G_0\choose l},\label{initial-condition} \\
Z(N,l,\mu,r=2)&=&0.
\label{initial-condition-r=2}
\end{eqnarray}
The second initial condition for $r=2$ follows from eq.~(\ref{qmax}).

We now introduce the functions $f_{k,d}$ depending on the integers $k$
and $d=0,1$
\begin{equation}
f_{k,d}(N,l,\mu,r)={N-l-k(r-2)-G_0(r+1)+d\choose
                N-l+1-k(r-1)-\mu -G_0(r+1)}
               {l+k(r-2)-d\choose
                l+k(r-1)-\mu}.
\end{equation}

\medskip\noindent{\bf\underbar{Lemma}: } {\em
The function
\begin{equation}
Z(N,l,\mu,r)=\sum_{k=-\infty}^\infty (f_{k,1}-f_{k,0})
\label{recurrence-solution}
\end{equation}
solves    the   recurrence  relation   (\ref{recurrence-relation}) and
fulfills the  initial condition (\ref{initial-condition}). Thus  it is
equal     to  the    partial     number   of    states    defined   in
eq.~(\ref{z-string-sum}). }

\medskip\noindent
The proof of this lemma is performed in Appendix C.

In    the   isotropic   limit    $r\to\infty$     all  the   functions
$f_{k,1}-f_{k,0}$ vanish  except that for  $k=0$ which  coincides with
the solution already found   by Bethe (1931).   One can  calculate the
total number of Bethe  ansatz states for  a given number of  roots $l$
(\ref{z-fixed-l-number}) by
\begin{equation}
Z(N,l,r) = \sum_{\mu =1}^l Z(N,l,\mu,r).
\end{equation}
For the interesting case $2j+1<r$ (with $j=N/2-l$) it has the form
\begin{equation}
Z(N,l,r)=\sum_{k=-\infty}^\infty\Gamma_{j+kr},\quad
\Gamma_{j+kr}=\sum_{\mu=1}^l(f_{k,1}-f_{k,0})
={N\choose l-kr}-{N\choose l-1-kr}
\end{equation}
which   indeed coincides  with    the    number of ``good''     path's
(\ref{good-path-number}).    Thus,  {\em  we   have  shown   that  the
  conjecture about Bethe ansatz states of the ``good'' type yields the
  correct number of states with respect  to the configuration space of
  the  quantum   invariant spin chain restricted    to the subspace of
  ``good'' representations.}

\section{Conclusions}
\label{sec:conclusions}

The configuration space of the quantum  group invariant XXZ Heisenberg
model  with open boundary conditions decomposes  into a ``good'' and a
``bad'' part, if and only  if the deformation  parameter $q$ is a root
of   unity.    Furthermore,       if    $q$   takes      the    values
\mbox{$q=\exp(\i\pi/r)$}    ($r=3,4,\dots$)  a   new  metric  may   be
introduced in the ``good'' subspace  such that the Hamiltonian becomes
selfadjoint.  This is reminiscent of minimal models of conformal field
theory.  The central charge  of  the Virasoro  algebra is known  to be
restricted by unitarity (Friedan et al 1984) to the values
\begin{equation}
  c=1-6/r(r-1),\quad r=3,4,5,\dots .
\end{equation}
Indeed, by finite  size computations (Hamer  et  al 1987) of  the spin
chain one obtains the identification to $c$.

In this paper  the completeness  of  the Bethe ansatz of  the  quantum
group symmetric  spin   $1/2$  Heisenberg  model  was  proved    for a
configuration space    reduced  to  the  ``good''  representations  at
$q=\exp(\i\pi /r)$   with $r=3,4,5,\dots$.  The ``good''  Bethe ansatz
solutions (in terms of  strings of positive parity) are  parameterized
by   integers (\ref{string-values})  which are  bounded  by the  upper
values $Q^{\max}_M$ (\ref{qmax}).   The conjecture for $Q^{\max}_M$ in
the quantum group case introduced   in this paper yields the   correct
number of states.    Furthermore, the ``good''  Bethe ansatz solutions
were checked numerically  for   small lattice length  ($N\leq  12$) by
solving the BAE exactly. In addition the eigenvalues obtained by these
solutions were compared with those  obtained by diagonalization of the
Hamiltonian in the path basis (see eq.~(\ref{temperly-lieb-action})).

\begin{appendix}
\section*{Appendix A: Non-string solutions of the BAE}
%\label{sec:bae-nonstring}

The Bethe ansatz  equations in the logarithmic form (\ref{bae-string})
provide a determination of solutions by a set of integers $\{Q_M\}$ if
one assumes   that  complex  roots    consist of   strings    $\lambda
=\Lambda\pm\i m$   with  known  imaginary  part  $m$.    However, this
assumption is known to  be only an approximation.   In general we have
complex pairs  $\lambda = x\pm\i y$ where  the  imaginary part is {\em
  not} a integer or  a half-integer (in  case of finite lattice length
$N$).   To consider  exact solutions  we analyze   the BAE (\ref{bae})
without this conjecture.  Unfortunately, for $l>2$ ($l\dots$ number of
roots) such   an  investigation can be   made  only  numerically.  But
analyzing a large  amount of numerical solutions we  were lead to some
rules which should be viewed as general properties of the Bethe ansatz
equations.

We have  found  (see appendix  B)  that by a  successive increasing of
$\gamma$ to integer  values $r=\pi  /\gamma$ ($q=\exp(\i\gamma)$)  the
largest admissible value  of $Q$ must  be  reduced in order to  obtain
finite rapidities of positive parity (which  are assumed to be related
to  ``good''  states).  Especially,   if  a 2-string ($l=2$)  tends to
infinity  at $Q$ and $r$   then a 2-string  at  $Q'=Q-1$ degenerate at
$r'=r-1$ into two 1-strings where one 1-string tends to infinity. This
leads to  the corrections $G_{1/2}$ and $G_{0}$  of the upper boundary
$Q^{\max}$.  In  the general case $l>2$,  we can  also assume that the
largest  value  $Q_M^{\max}$  for a    set of  integers  is caused  by
non-string effects.   By numerical computations   we have observed the
following important property.
\begin{quote}
  If a k-string (string  of length k=2M+1)  tends to infinity at $Q_M$
  and $r$   then a k-string at   $Q_M'=Q_M-1$ and $r'=r-1$ degenerates
  into a divergent (k-1)-string and  a 1-string.  Moreover, a k-string
  at  $Q_M''=Q_M-2$   and  $r''=r-2$   degenerates  into   a divergent
  (k-2)-string and a 2-string.  In general, a k-string at $Q_M'=Q_M-n$
  and  $r'=r-n$ degenerates and the  new  (k-n)-string or the n-string
  becomes infinite.
\end{quote}
Due to this conjecture we obtain the condition
\begin{equation}
G_M(r)=G_M(r-1)-1.
\end{equation}
Considering the value $Q^\infty_M$ in equation (\ref{string-infty})
$$ Q_M^{\infty} = N-2\sum_{M'}\nu_{M'}J(M,M')+
   (2M+1)(1-\frac{\gamma}{\pi}(N-2l+2M+2))
$$
one can assume that the initial value $G_M(r)=1$
is given by
\begin{equation}
\frac{\gamma}{\pi}(N-2l+2M+2)=1
\end{equation}
which leads to
\begin{equation}
G_M(r)=2j+(2M+1)-r+2
\end{equation}
as was suggested in (\ref{qmax}).

We point out that this picture of admissible strings holds only in the
case  $2j+1<r$. For $2j+1\geq  r$  there are more configurations which
leads to  finite  solutions  of positive parity.   Nevertheless,  such
states always belong to the ``bad'' type and have to be excluded.

\section*{Appendix B: Numerical computation of the BAE}
%\label{sec:numerical-results}

Considering  the   two-magnon case  $l=2$   one obtains  two kinds  of
possible solutions.  Due to (\ref{bae})  there are either two distinct
real  roots  $\{\Lambda_{1},\Lambda_{2}\}$  or  a complex pair $\{x+\i
y,x-\i y\}$.  The  two real roots  satisfy the following Bethe  ansatz
equations
\begin{eqnarray}
2\pi Q_1&=& 2N\Psi_{1/2}(\Lambda_1)-
           \Psi_1(\Lambda_1-\Lambda_2)-\Psi_1(\Lambda_1+\Lambda_2) \\
2\pi Q_2&=& 2N\Psi_{1/2}(\Lambda_2)-
           \Psi_1(\Lambda_2-\Lambda_1)-\Psi_1(\Lambda_2+\Lambda_1)
\label{bae-two-magnon-real}
\end{eqnarray}
corresponding to two 1-strings. Because of the zero imaginary part the
solution is exactly  given  by (\ref{bae-string}) with  $\nu_0=2$  and
$\nu_k=0$ ($k>0$). If one root tends to infinity the value $Q$ reads
\begin{equation}
Q^\infty =2j+2-\frac{\gamma}{\pi}(2j+2).
\end{equation}
Thus,  for $(2j+2)\pi  /\gamma<1$ the  largest  admissible integer  is
given by
\begin{equation}
Q^{\max}_0=2j+2
\end{equation}
and  for  $\pi /\gamma =r=(2j+2)$   one  root tends to  infinity.  The
condition
\begin{equation}
Q^{\max}_0=2j+2-G_0,\quad G_0=1=2j+3-r
\end{equation}
ensures finite  solutions. For example,  the case $N=8$ ($l=2$, $j=2$)
and  $r>2j+2$ provides configurations  of integers $\{Q_1,Q_2\}$ which
are in one-to-one correspondence to the solutions of the BAE
$$ \{\Lambda_1,\Lambda_2\} \leftrightarrow  \{Q_1,Q_2\} $$
with $1\leq Q_1<Q_2\leq Q^{\max}_0=2j+2=6$.  At $r=2j+2$ the number of
configurations is reduced by $Q^{\max}_0=6-1$. For $r<2j+2$ we have no
``good'' state.

The  case of complex  roots ($l=2$)  is   more complicated. Using  the
string conjecture we would have the following equation for the complex
pair (\ref{bae-string})
\begin{equation}
\pi Q_{1/2}=N\Psi_1 (\Lambda_{1/2})-\Psi_1 (2\Lambda_{1/2})
\label{bae-string-two-magnon}
\end{equation}
with the configuration $\nu_{1/2}=1,\quad   \nu_{0}=0,\nu_{1}=0\dots$.
As a function of $\Lambda$ the r.h.s. is monotone and all integers $Q$
up to the asymptotic value
$$ Q_{1/2}^\infty =2j+\nu_{1/2}+2-2(2j+3)/r $$
would lead to  a finite value  $\Lambda_{1/2}$.   This contradicts the
condition that the largest admissible integer is given by
\begin{equation}
Q_{1/2}^{\max}=2j+\nu_{1/2}-G_{1/2}=2j+1-G_{1/2}
\end{equation}
Even in   the XXX  case  ($G_{1/2}=0$)  there are integers   $Q$  with
$Q_{1/2}^{\max}<Q<Q_{1/2}^{\infty}$    which     are    allowed     in
(\ref{bae-string-two-magnon})  but  forbidden  as Bethe ansatz states.
We  have  already  pointed out that   for such  large  values $Q$  the
imaginary part of the root tends  away from the assumed (half) integer
value  and therefore the string conjecture  does not hold shown by the
exact solutions of the BAE.  A complex root  denoted by $\lambda =x+\i
y$ satisfies the equations (\ref{bae})
\begin{equation}
\left(\frac{\sinh\gamma (x+\i (\pm y+1/2))}
           {\sinh\gamma (x+\i (\pm y-1/2))}
\right)^{2N} =
\frac{\sin\gamma (\pm 2y+1 )}
     {\sin\gamma (\pm 2y-1)}
\frac{\sinh\gamma (2x+\i )}
     {\sinh\gamma (2x-\i )}.
\label{bae-two-magnon}
\end{equation}
Considering the phase of these relations
\begin{eqnarray}
\pi Q_{1/2} &=& -N\phi_{1/2\pm y}(x)-N\phi_{1/2\mp y}(x)+
          \phi_1(2x)+\pi (N-1)
\label{bae-two-magnon-phase}\\
\phi_y (x)&=& 2\arctan (\tan\gamma y\coth\gamma x)
\end{eqnarray}
we obtain in the  limit $y\to  1/2$ (with $\phi_y(x)=\pi  -\Psi_y(x)$)
the  equation for  a two-string (\ref{bae-string-two-magnon}).   Thus,
these integers $Q$  are related to (\ref{bae-string-two-magnon}).  The
magnitude reads
\begin{equation}
\left(\frac{\sinh^2\gamma x+\sinh^2\gamma (\pm y+1/2))}
           {\sinh^2\gamma x+\sinh^2\gamma (\pm y-1/2))}
\right)^{2N} =
\frac{\sin^2\gamma (\pm 2y+1 )}
     {\sin^2\gamma (\pm 2y-1)}
\end{equation}
which is used to eliminate the real part $x$ in
(\ref{bae-two-magnon-phase}). With
\begin{eqnarray}
x &=& b(y) \\
\sinh^2\gamma b(y) &=&
   \frac{W_2(y)^{1/2N}-W_1(y)}{W_1(y)(1-W_2(y)^{1/2N})}
   \sinh^2\gamma(y+1/2) \\
W_m(y)&=&\frac{\sinh^2\gamma m(y+1/2)}{\sinh^2\gamma m(y-1/2)}
\end{eqnarray}
the counting function $z(y)$ depending on $y$ reads
\begin{equation}
\pi z(y)= -N\left(\phi_{1/2+y}(b(y))+\phi_{1/2-y}(b(y))\right)+
          \phi_1(2b(y))+\pi (N-1).
\label{z-function-two-magnon}
\end{equation}
It turns out that a solution for $\mid y\mid <1/2$ and
$\mid y\mid >1/2$ is given by even and odd integers, respectively
\begin{equation}
Q_{1/2}=z(y).
\end{equation}
\ifpictex
\input prepictex \input pictex \input postpictex
\typeout{Create PiCTeX figure. Please wait... }
\begin{figure}[hbt]
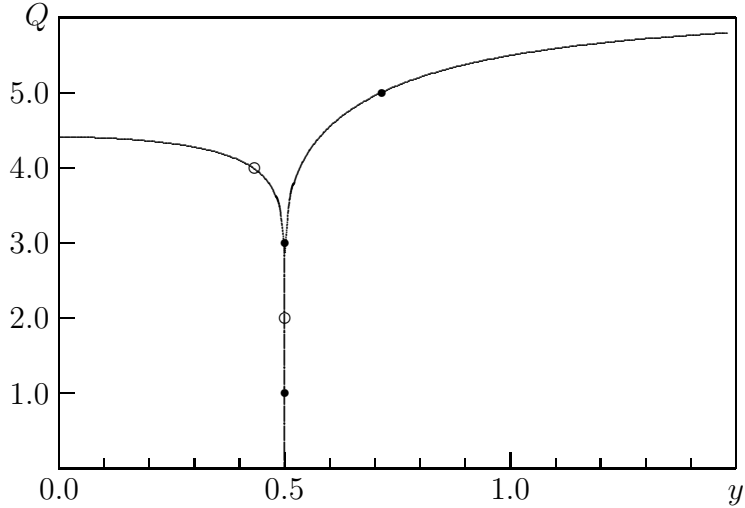

\[ \beginpicture
\setcoordinatesystem units <6cm,1cm>
\setplotarea x from 0.0 to 1.5, y from 0.0 to 6.0
\axis bottom label {}  ticks in
        numbered from 0.0 to 1.0 by 0.5
        withvalues $y$ / at 1.5 /
        unlabeled short from 0.1 to 1.4 by 0.1 /
\axis left   label {} ticks in
                   numbered from 1.0 to 5.0 by 1.0
                   withvalues $Q$ / at 6.0 /
                   unlabeled short from 1.0 to 5.0 by 1.0 /
\axis right  label {}  /
\axis top    label {}  /
\iffinalplot
\setquadratic
\plot 0.001   4.40754
0.011   4.4074
0.021   4.40703
0.031   4.40642
0.041   4.40558
0.051   4.40451
0.061   4.40319
0.071   4.40163
0.081   4.39983
0.091   4.39777
0.101   4.39545
0.111   4.39287
0.121   4.39001
0.131   4.38687
0.141   4.38344
0.151   4.37971
0.161   4.37567
0.171   4.3713
0.181   4.36658
0.191   4.36151
0.201   4.35606
0.211   4.35021
0.221   4.34394
0.231   4.33723
0.241   4.33003
0.251   4.32232
0.261   4.31407
0.271   4.30523
0.281   4.29574
0.291   4.28557
0.301   4.27463
0.311   4.26286
0.321   4.25018
0.331   4.23647
0.341   4.22163
0.351   4.20551
0.361   4.18793
0.371   4.16869
0.381   4.14753
0.391   4.12412
0.401   4.09803
0.411   4.06871
0.421   4.03542
0.431   3.9971
0.441   3.95225
0.451   3.8985
0.461   3.83195
0.471   3.7453
0.481   3.62223
0.491   3.41011
0.501   2.86386
0.511   3.54998
0.521   3.79245
0.531   3.9562
0.541   4.08319
0.551   4.18823
0.561   4.27837
0.571   4.35763
0.581   4.42851
0.591   4.49267
0.601   4.55133
0.611   4.60536
0.621   4.65543
0.631   4.70207
0.641   4.74571
0.651   4.78668
0.661   4.82527
0.671   4.86172
0.681   4.89623
0.691   4.92898
0.701   4.96011
0.711   4.98977
0.721   5.01806
0.731   5.04508
0.741   5.07094
0.751   5.0957
0.761   5.11944
0.771   5.14223
0.781   5.16413
0.791   5.18518
0.801   5.20545
0.811   5.22497
0.821   5.2438
0.831   5.26195
0.841   5.27948
0.851   5.29641
0.861   5.31277
0.871   5.32859
0.881   5.3439
0.891   5.35873
0.901   5.37308
0.911   5.387
0.921   5.40049
0.931   5.41357
0.941   5.42626
0.951   5.43859
0.961   5.45055
0.971   5.46218
0.981   5.47348
0.991   5.48446
1.001   5.49514
1.011   5.50553
1.021   5.51564
1.031   5.52548
1.041   5.53506
1.051   5.5444
1.061   5.55349
1.071   5.56235
1.081   5.57098
1.091   5.5794
1.101   5.58761
1.111   5.59562
1.121   5.60344
1.131   5.61106
1.141   5.61851
1.151   5.62577
1.161   5.63287
1.171   5.6398
1.181   5.64658
1.191   5.65319
1.201   5.65966
1.211   5.66598
1.221   5.67216
1.231   5.6782
1.241   5.68411
1.251   5.68989
1.261   5.69555
1.271   5.70108
1.281   5.7065
1.291   5.7118
1.301   5.71699
1.311   5.72207
1.321   5.72704
1.331   5.73192
1.341   5.73669
1.351   5.74137
1.361   5.74595
1.371   5.75044
1.381   5.75483
1.391   5.75915
1.401   5.76337
1.411   5.76752
1.421   5.77158
1.431   5.77557
1.441   5.77947
1.451   5.78331
1.461   5.78707
1.471   5.79076
1.481   5.79438
/
\plot 0.5 3.0  0.5 1.5  0.5 0.0 /
\multiput {\circle*{3}} [Bl] at 0.715 5.0
                                0.5   3.0
                                0.5   1.0 /
\multiput {\circle{4}} [Bl] at  0.433 4.0
                                0.5   2.0 /
\fi
\endpicture \]
\caption{Counting function for a complex pair $\lambda =x\pm\i y$ with
         $N=8$, $l=2$ and $r=14$ {\em as a function of $y$}.
         Full (open) circles correspond to odd (even) values of $Q$.}
\label{fig:z-function}
\vspace{\topskip}
\end{figure}
\typeout{PiCTeX figure. Done.}
\fi
The general behavior of $z$ drawn in figure \ref{fig:z-function}
shows that for small values of $Q$ the imaginary part $y$ of a complex
pair is approximated very well by $y=1/2$ with the correspondence
$$ \{x+\i y,x-\i y\}\leftrightarrow\{Q_{1/2}\}. $$
But for large values $Q$ the
imaginary part of the root tends away from the line $y=1/2$.
Now we discuss the example $N=8$ but it can be easily generalized. The
counting function $z(y)$ takes it maximum at $y=0$ (for $y<1/2$) and at
$y=\pi /4\gamma =r/4$.
One can show
that for $\gamma\to 0$ (XXX case) the maximal integer is determined by
$Q_{1/2}^{\max}=2j+1$ which corresponds to the upper
restriction stated above.
Although the maximum $z^{\max}$ (at $y=r/4$) is larger than $6$
$$ 6<z^{\max}<7,\quad y>1/2 $$
the largest admissible integer is related to
$$ Q_{1/2}^{\max}=2j+1=5. $$
Note that the odd integers $Q_{1/2}=1,3,5$ determine
the solutions with $y>1/2$.
The even integers $Q_{1/2}=2,4$ correspond to roots with $y<1/2$ because of
$$ 4<z^{\max}<5,\quad y<1/2. $$
The maximum of $z(y)$ for both $y<1/2$ and $y>1/2$ depends on the
anisotropy $\gamma$. It
decreases with increasing $\gamma$.
The picture described above remains unchanged up to values
$r>2j+3$. In the case
$r=8$ for example we have $z^{\max}_{y>1/2}=5.25$ and
$z^{\max}_{y<1/2}=4.07195\dots$. Thus, the integers
$Q_{1/2}=1,2,3,4,5$ are
assumed to correspond to ``good'' state if $r\geq 8$.

Increasing $\gamma$ we have at
$$ \pi /\gamma =r=2j+3=7 $$
(according to (\ref{z-function-two-magnon}))
a divergent complex pair with $Q_{1/2}=2j+1=5$
($x\to\infty$, $y=\pm r/4$) because of $z^{\max}_{y>1/2}=5$.
Therefore, the value $Q_{1/2}^{\max}$ for finite roots must be reduced
by one
$$ Q^{\max}_{1/2}=2j+1-1=4 $$
which leads to $G_{1/2}=1$.
The odd value $Q_{1/2}=2j+1=5$ is now forbidden.
Furthermore, by increasing $\gamma$ {\em or} increasing $N$ we
reach a critical value where the maximum of $z(y)$ for $y<1/2$
is smaller than the even integer $Q_{1/2}=2j=4$
$$ z^{\max}_{y<1/2}=3.93162\dots<2j, \quad r=7.  $$ The imaginary part
$y$ tends to  zero. This fact  was observed  by  Vladimirov (1984) and
E\ss ler et al (1991b) in the XXX case. The complex pair was found to be
replaced  by  two additional   {\em real}  roots.  The  dependence  on
$\gamma$ was investigated  by J\"uttner et al  (1993).  Here we have a
similar  situation.  The additional  real solution is  considered as a
degenerate complex pair corresponding to $Q_{1/2}=2j=4$. The roots are
finite and therefore remain  ``good''.  Such a degenerate solution  is
described   by   (\ref{bae-two-magnon-real})  having  {\em  identical}
integers
$$ Q_1=Q_2=Q_0^{\max}=2j+2=6 $$
at $Q_0^{\max}$ although their rapidities are distinct
$\Lambda_1\neq\Lambda_2$. This correspondence is denoted by
$$
\{\Lambda_1,\Lambda_2\}
\leftrightarrow\{Q_0^{\max},Q_0^{\max}\}
\leftrightarrow\{Q_{1/2}\}.
$$
Note that there is a difference between $Q_0^{\max}$ of real roots and
the value $Q_{1/2}$ for a 2-string. At $r=7$ we consider the values
$Q_{1/2}=1,2,3,4$ as ``good''  integers.

Increasing $\gamma$ to $r=6$ we will see that a solution related to
$Q_{1/2}=4$ becomes ``bad''.
It can be shown
that in the limit $\pi /\gamma \to 2j+2$ a complex pair at $Q_{1/2}=2j=4$
always degenerate.
But at $r=2j+2=6$ a real root related to
$Q=2j+2=Q_0^{\max}=6$ is known to tend to infinity. This happens also for a
degenerate complex pair.
Because one part of this degenerate string is divergent we must exclude
the corresponding integer {\em of the 2-string}
$$ Q_{1/2}^{\max}=2j-1 $$
which leads to $G_{1/2}=2$ and ``good'' values $Q_{1/2}=1,2,3$.

A further increasing of $\gamma$ to $r\leq 5$ leads to the case where
any state belongs to the ``bad'' sector. Namely, the total spin
($j=2$) is related to $2j+1\geq r$. One can check that at $r=5$ the
admissible numbers for finite solutions of positive parity read
$Q_{1/2}=1,2,3$. Thus we have no further reduction of
$Q_{1/2}^{\max}$. But this is of no interest for our consideration of
``good'' states.

The case $l>2$ is treated as follows. Fixing a given set of integers
$\{Q_M\}$ we first calculate the centers of strings
$\{\Lambda_M\}_{\gamma_0}$ for $0<\gamma_0 <<1$ by fixed point iterations of
the BAE (\ref{bae-string})
\begin{equation}
\Lambda_{M,i} = \frac{1}{\gamma}\tanh^{-1}\left(\tan\gamma m
\tan\left(2\pi Q_{M,i} +
\sum_{m=1}^{2M}2\Psi_m(2\Lambda_{M,i})+
\sum_{M'=0}\sum_{k=1}^{\nu_{M'}}\dots
\right) /4N \right)
\end{equation}
It turns out that the iteration converges (nearly) independently
on the initial values for $\Lambda_{M,i}$. The set of centers
$\{\Lambda_M\}_{\gamma_0}$ is now a function of $\{Q_M\}$.
The next step consists in an exact computation of the roots
$\{\lambda\}$ by the original BAE (\ref{bae}) which is numerically
solved by the Newton method. As initial values we use the string
centers $\{\Lambda_M\}_{\gamma_0}$ where the corresponding imaginary part
of each member reads
$$ \lambda =\Lambda_M +\i m+\i\delta $$
with $0<\mid\delta\mid <<1$. With a careful choice of $\delta$ the
Newton method converges which yields the accurate solutions of the BAE
for $\gamma_0$. Thus, we have a correspondence between the set $\{Q_M\}$
and $\{\lambda\}$
$$ \{\lambda\}_{\gamma_0}=f(\{Q_M\},\gamma_0). $$
Now the anisotropy $\gamma$ is increased in small steps
$$ \gamma_{k+1}=\gamma_k+\Delta . $$
For each $\gamma_k$ the Newton method is applied. In contrast to the
first step we now take initial values which are accurate
solutions in the step $k-1$. If $\Delta$ is small enough one can
assume that the roots $\{\lambda\}_{\gamma_k}$ keep the relation to
$\{\lambda\}_{\gamma_{k-1}}$ with respect to the correspondence to the
integers  $\{Q_M\}$. This provides
$$  \{\lambda\}_{\gamma_k}=f(\{Q_M\},\gamma_k) $$
which can be used to investigate the behavior of the solutions as a
function of the integers $\{Q_M\}$.

Now we discuss some numerical examples. Considering a 3-string for
$N=8$ and $l=3$ ($\nu_1=1$) the integers $Q=1,2,3$ are allowed
($\pi /\gamma >6$)
and the exact solutions are listed in table \ref{tab:3-string}.
\begin{table}[tb]
\caption{Exact solutions for $N=8,\quad l=3$.}
\label{tab:3-string}
{\footnotesize
$$
\renewcommand{\arraystretch}{1}
\begin{array}{|c||c|c||c|c||c|c|}\hline
 &\multicolumn{2}{|c||}{Q=1}
 &\multicolumn{2}{|c||}{Q=2}
 &\multicolumn{2}{|c|}{Q=3} \\ \hline
\pi /\gamma &\lambda_1&\lambda_{2,3}
            &\lambda_1&\lambda_{2,3}
            &\lambda_1&\lambda_{2,3} \\ \hline
7& 0.672828 & 0.672829\pm i 1.000377
 & 1.305862 & 1.338141\pm i 1.035500
 & 2.379542 & 2.183152\pm i 1.348160 \\
6& 0.716146 & 0.716398\pm i 1.000649
 & 1.407535 & 1.489400\pm i 1.060994
 & \infty   & \infty  \pm i \pi /3\gamma \\
5& 0.810005 & 0.811995\pm i 1.001089
 & 1.681845 & \infty  \pm i \pi /4\gamma
 && \\
4& \infty   & 1.376812\pm i 0.569724
 && &&\\ \hline
\end{array}
$$
}
\end{table}
The numbers $Q=1,2,3$ are associated to the following
picture where the real and imaginary part of the roots are plotted in
the x-y-plane. At $\gamma =\pi /7$ we obtain
$$
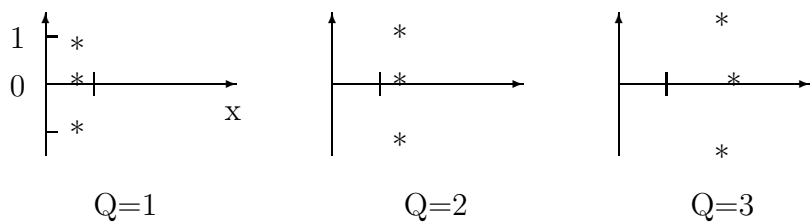
\setlength{\unitlength}{0.012500in}%
\begingroup\makeatletter\ifx\SetFigFont\undefined
% extract first six characters in \fmtname
\def\x#1#2#3#4#5#6#7\relax{\def\x{#1#2#3#4#5#6}}%
\expandafter\x\fmtname xxxxxx\relax \def\y{splain}%
\ifx\x\y   % LaTeX or SliTeX?
\gdef\SetFigFont#1#2#3{%
  \ifnum #1<17\tiny\else \ifnum #1<20\small\else
  \ifnum #1<24\normalsize\else \ifnum #1<29\large\else
  \ifnum #1<34\Large\else \ifnum #1<41\LARGE\else
     \huge\fi\fi\fi\fi\fi\fi
  \csname #3\endcsname}%
\else
\gdef\SetFigFont#1#2#3{\begingroup
  \count@#1\relax \ifnum 25<\count@\count@25\fi
  \def\x{\endgroup\@setsize\SetFigFont{#2pt}}%
  \expandafter\x
    \csname \romannumeral\the\count@ pt\expandafter\endcsname
    \csname @\romannumeral\the\count@ pt\endcsname
  \csname #3\endcsname}%
\fi
\fi\endgroup
\begin{picture}(335,90)(5,745)
\thinlines
\put(140,770){\vector( 0, 1){ 60}}
\put(260,770){\vector( 0, 1){ 60}}
\put(260,800){\vector( 1, 0){ 80}}
\put( 20,820){\line( 1, 0){  5}}
\put( 20,780){\line( 1, 0){  5}}
\put( 40,805){\line( 0,-1){ 10}}
\put(160,805){\line( 0,-1){ 10}}
\put(280,805){\line( 0,-1){ 10}}
\put( 20,800){\vector( 1, 0){ 80}}
\put(140,800){\vector( 1, 0){ 80}}
\put(  5,815){\makebox(0,0)[lb]{\smash{\SetFigFont{12}{14.4}{rm}1}}}
\put(  5,795){\makebox(0,0)[lb]{\smash{\SetFigFont{12}{14.4}{rm}0}}}
\put( 20,770){\vector( 0, 1){ 60}}
\put( 95,785){\makebox(0,0)[lb]{\smash{\SetFigFont{12}{14.4}{rm}x}}}
\put(305,795){\makebox(0,0)[lb]{\smash{\SetFigFont{12}{14.4}{rm}*}}}
\put(165,815){\makebox(0,0)[lb]{\smash{\SetFigFont{12}{14.4}{rm}*}}}
\put(165,770){\makebox(0,0)[lb]{\smash{\SetFigFont{12}{14.4}{rm}*}}}
\put(300,765){\makebox(0,0)[lb]{\smash{\SetFigFont{12}{14.4}{rm}*}}}
\put(300,820){\makebox(0,0)[lb]{\smash{\SetFigFont{12}{14.4}{rm}*}}}
\put( 30,775){\makebox(0,0)[lb]{\smash{\SetFigFont{12}{14.4}{rm}*}}}
\put( 30,810){\makebox(0,0)[lb]{\smash{\SetFigFont{12}{14.4}{rm}*}}}
\put( 40,745){\makebox(0,0)[lb]{\smash{\SetFigFont{12}{14.4}{rm}Q=1}}}
\put(170,745){\makebox(0,0)[lb]{\smash{\SetFigFont{12}{14.4}{rm}Q=2}}}
\put(290,745){\makebox(0,0)[lb]{\smash{\SetFigFont{12}{14.4}{rm}Q=3}}}
\put( 30,795){\makebox(0,0)[lb]{\smash{\SetFigFont{12}{14.4}{rm}*}}}
\put(165,795){\makebox(0,0)[lb]{\smash{\SetFigFont{12}{14.4}{rm}*}}}
\end{picture}
$$
The roots are arranged as 3-strings very well. Increasing
$\gamma$ to $\pi /6$ the roots related to $Q=3$ tend to infinity - the
real root as well as the real part of the complex pair.
$$
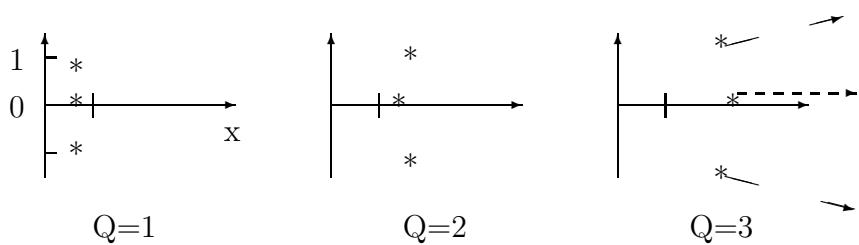
\setlength{\unitlength}{0.012500in}%
\begingroup\makeatletter\ifx\SetFigFont\undefined
% extract first six characters in \fmtname
\def\x#1#2#3#4#5#6#7\relax{\def\x{#1#2#3#4#5#6}}%
\expandafter\x\fmtname xxxxxx\relax \def\y{splain}%
\ifx\x\y   % LaTeX or SliTeX?
\gdef\SetFigFont#1#2#3{%
  \ifnum #1<17\tiny\else \ifnum #1<20\small\else
  \ifnum #1<24\normalsize\else \ifnum #1<29\large\else
  \ifnum #1<34\Large\else \ifnum #1<41\LARGE\else
     \huge\fi\fi\fi\fi\fi\fi
  \csname #3\endcsname}%
\else
\gdef\SetFigFont#1#2#3{\begingroup
  \count@#1\relax \ifnum 25<\count@\count@25\fi
  \def\x{\endgroup\@setsize\SetFigFont{#2pt}}%
  \expandafter\x
    \csname \romannumeral\the\count@ pt\expandafter\endcsname
    \csname @\romannumeral\the\count@ pt\endcsname
  \csname #3\endcsname}%
\fi
\fi\endgroup
\begin{picture}(355,90)(5,745)
\thinlines
\put(140,770){\vector( 0, 1){ 60}}
\put(260,770){\vector( 0, 1){ 60}}
\put(260,800){\vector( 1, 0){ 80}}
\put( 20,820){\line( 1, 0){  5}}
\put( 20,780){\line( 1, 0){  5}}
\put( 40,805){\line( 0,-1){ 10}}
\put(160,805){\line( 0,-1){ 10}}
\put(280,805){\line( 0,-1){ 10}}
\put( 20,800){\vector( 1, 0){ 80}}
\multiput(305,770)(40.10443,-10.02611){2}{\line( 4,-1){ 14.013}}
\put(359,756){\vector( 4,-1){0}}
\multiput(310,805)(7.69231,0.00000){7}{\line( 1, 0){  3.846}}
\put(360,805){\vector( 1, 0){0}}
\multiput(305,825)(35.39863,8.84966){2}{\line( 4, 1){ 14.013}}
\put(354,837){\vector( 4, 1){0}}
\put(140,800){\vector( 1, 0){ 80}}
\put( 20,770){\vector( 0, 1){ 60}}
\put(  5,815){\makebox(0,0)[lb]{\smash{\SetFigFont{12}{14.4}{rm}1}}}
\put(170,770){\makebox(0,0)[lb]{\smash{\SetFigFont{12}{14.4}{rm}*}}}
\put(  5,795){\makebox(0,0)[lb]{\smash{\SetFigFont{12}{14.4}{rm}0}}}
\put( 95,785){\makebox(0,0)[lb]{\smash{\SetFigFont{12}{14.4}{rm}x}}}
\put(300,765){\makebox(0,0)[lb]{\smash{\SetFigFont{12}{14.4}{rm}*}}}
\put(300,820){\makebox(0,0)[lb]{\smash{\SetFigFont{12}{14.4}{rm}*}}}
\put( 30,775){\makebox(0,0)[lb]{\smash{\SetFigFont{12}{14.4}{rm}*}}}
\put( 30,810){\makebox(0,0)[lb]{\smash{\SetFigFont{12}{14.4}{rm}*}}}
\put( 40,745){\makebox(0,0)[lb]{\smash{\SetFigFont{12}{14.4}{rm}Q=1}}}
\put(170,745){\makebox(0,0)[lb]{\smash{\SetFigFont{12}{14.4}{rm}Q=2}}}
\put(290,745){\makebox(0,0)[lb]{\smash{\SetFigFont{12}{14.4}{rm}Q=3}}}
\put( 30,795){\makebox(0,0)[lb]{\smash{\SetFigFont{12}{14.4}{rm}*}}}
\put(305,795){\makebox(0,0)[lb]{\smash{\SetFigFont{12}{14.4}{rm}*}}}
\put(165,795){\makebox(0,0)[lb]{\smash{\SetFigFont{12}{14.4}{rm}*}}}
\put(170,815){\makebox(0,0)[lb]{\smash{\SetFigFont{12}{14.4}{rm}*}}}
\end{picture}
$$
Thus, the admissible numbers are reduced to $Q=1,2$. Now, in the limit
$\gamma\to\pi /5$ the string at $Q=2$ degenerate. The real root
remains finite - but the difference to the real part of the complex
pair becomes larger. This 3-string degenerate into a 2-string and a
1-string in such a manner that at $\pi /5$ the complex pair tends to
infinity.
$$
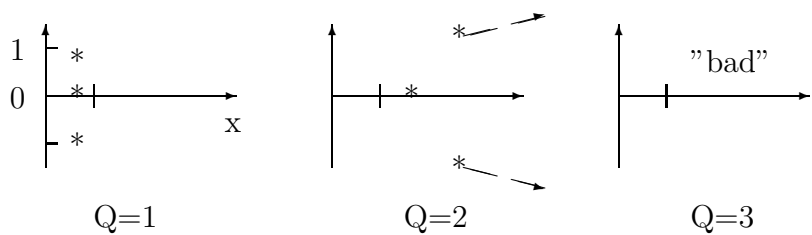
\setlength{\unitlength}{0.012500in}%
\begingroup\makeatletter\ifx\SetFigFont\undefined
% extract first six characters in \fmtname
\def\x#1#2#3#4#5#6#7\relax{\def\x{#1#2#3#4#5#6}}%
\expandafter\x\fmtname xxxxxx\relax \def\y{splain}%
\ifx\x\y   % LaTeX or SliTeX?
\gdef\SetFigFont#1#2#3{%
  \ifnum #1<17\tiny\else \ifnum #1<20\small\else
  \ifnum #1<24\normalsize\else \ifnum #1<29\large\else
  \ifnum #1<34\Large\else \ifnum #1<41\LARGE\else
     \huge\fi\fi\fi\fi\fi\fi
  \csname #3\endcsname}%
\else
\gdef\SetFigFont#1#2#3{\begingroup
  \count@#1\relax \ifnum 25<\count@\count@25\fi
  \def\x{\endgroup\@setsize\SetFigFont{#2pt}}%
  \expandafter\x
    \csname \romannumeral\the\count@ pt\expandafter\endcsname
    \csname @\romannumeral\the\count@ pt\endcsname
  \csname #3\endcsname}%
\fi
\fi\endgroup
\begin{picture}(335,90)(5,745)
\thinlines
\put(140,770){\vector( 0, 1){ 60}}
\put(260,770){\vector( 0, 1){ 60}}
\put(260,800){\vector( 1, 0){ 80}}
\put( 20,820){\line( 1, 0){  5}}
\put( 20,780){\line( 1, 0){  5}}
\put( 40,805){\line( 0,-1){ 10}}
\put(160,805){\line( 0,-1){ 10}}
\put(280,805){\line( 0,-1){ 10}}
\put( 20,800){\vector( 1, 0){ 80}}
\put(140,800){\vector( 1, 0){ 80}}
\multiput(195,770)(20.10443,-5.02611){2}{\line( 4,-1){ 14.013}}
\put(229,761){\vector( 4,-1){0}}
\multiput(195,825)(20.10443,5.02611){2}{\line( 4, 1){ 14.013}}
\put(229,834){\vector( 4, 1){0}}
\put( 20,770){\vector( 0, 1){ 60}}
\put(  5,815){\makebox(0,0)[lb]{\smash{\SetFigFont{12}{14.4}{rm}1}}}
\put(190,765){\makebox(0,0)[lb]{\smash{\SetFigFont{12}{14.4}{rm}*}}}
\put(  5,795){\makebox(0,0)[lb]{\smash{\SetFigFont{12}{14.4}{rm}0}}}
\put( 95,785){\makebox(0,0)[lb]{\smash{\SetFigFont{12}{14.4}{rm}x}}}
\put( 30,775){\makebox(0,0)[lb]{\smash{\SetFigFont{12}{14.4}{rm}*}}}
\put( 30,810){\makebox(0,0)[lb]{\smash{\SetFigFont{12}{14.4}{rm}*}}}
\put( 40,745){\makebox(0,0)[lb]{\smash{\SetFigFont{12}{14.4}{rm}Q=1}}}
\put(170,745){\makebox(0,0)[lb]{\smash{\SetFigFont{12}{14.4}{rm}Q=2}}}
\put(290,745){\makebox(0,0)[lb]{\smash{\SetFigFont{12}{14.4}{rm}Q=3}}}
\put( 30,795){\makebox(0,0)[lb]{\smash{\SetFigFont{12}{14.4}{rm}*}}}
\put(290,810){\makebox(0,0)[lb]{\smash{\SetFigFont{12}{14.4}{rm}"bad"}}}
\put(170,795){\makebox(0,0)[lb]{\smash{\SetFigFont{12}{14.4}{rm}*}}}
\put(190,820){\makebox(0,0)[lb]{\smash{\SetFigFont{12}{14.4}{rm}*}}}
\end{picture}
$$
Now we have only one allowed integer $Q=1$.

The behavior of a 4-string is similar. If $N=10$ and $l=4$,
$\nu_{3/2}=1$ we have 3 admissible values $Q=1,2,3$
($\gamma <\pi/7$).
$$
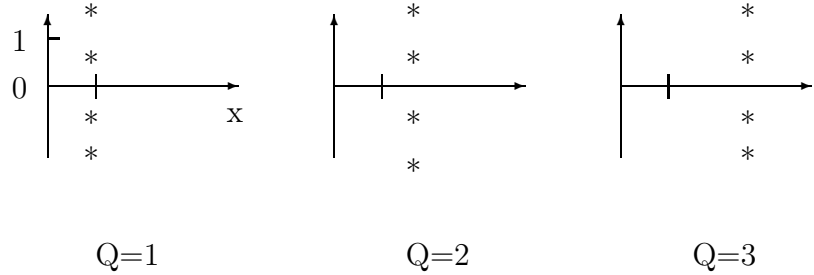
\setlength{\unitlength}{0.012500in}%
\begingroup\makeatletter\ifx\SetFigFont\undefined
% extract first six characters in \fmtname
\def\x#1#2#3#4#5#6#7\relax{\def\x{#1#2#3#4#5#6}}%
\expandafter\x\fmtname xxxxxx\relax \def\y{splain}%
\ifx\x\y   % LaTeX or SliTeX?
\gdef\SetFigFont#1#2#3{%
  \ifnum #1<17\tiny\else \ifnum #1<20\small\else
  \ifnum #1<24\normalsize\else \ifnum #1<29\large\else
  \ifnum #1<34\Large\else \ifnum #1<41\LARGE\else
     \huge\fi\fi\fi\fi\fi\fi
  \csname #3\endcsname}%
\else
\gdef\SetFigFont#1#2#3{\begingroup
  \count@#1\relax \ifnum 25<\count@\count@25\fi
  \def\x{\endgroup\@setsize\SetFigFont{#2pt}}%
  \expandafter\x
    \csname \romannumeral\the\count@ pt\expandafter\endcsname
    \csname @\romannumeral\the\count@ pt\endcsname
  \csname #3\endcsname}%
\fi
\fi\endgroup
\begin{picture}(335,115)(5,705)
\thinlines
\put( 20,750){\vector( 0, 1){ 60}}
\put( 20,780){\vector( 1, 0){ 80}}
\put( 20,800){\line( 1, 0){  5}}
\put( 40,785){\line( 0,-1){ 10}}
\put(140,750){\vector( 0, 1){ 60}}
\put(140,780){\vector( 1, 0){ 80}}
\put(160,785){\line( 0,-1){ 10}}
\put(260,750){\vector( 0, 1){ 60}}
\put(260,780){\vector( 1, 0){ 80}}
\put(280,785){\line( 0,-1){ 10}}
\put( 40,705){\makebox(0,0)[lb]{\smash{\SetFigFont{12}{14.4}{rm}Q=1}}}
\put( 95,765){\makebox(0,0)[lb]{\smash{\SetFigFont{12}{14.4}{rm}x}}}
\put(  5,775){\makebox(0,0)[lb]{\smash{\SetFigFont{12}{14.4}{rm}0}}}
\put( 20,780){\line( 1, 0){  5}}
\put(  5,795){\makebox(0,0)[lb]{\smash{\SetFigFont{12}{14.4}{rm}1}}}
\put(310,745){\makebox(0,0)[lb]{\smash{\SetFigFont{12}{14.4}{rm}*}}}
\put( 35,760){\makebox(0,0)[lb]{\smash{\SetFigFont{12}{14.4}{rm}*}}}
\put( 35,785){\makebox(0,0)[lb]{\smash{\SetFigFont{12}{14.4}{rm}*}}}
\put( 35,805){\makebox(0,0)[lb]{\smash{\SetFigFont{12}{14.4}{rm}*}}}
\put( 35,745){\makebox(0,0)[lb]{\smash{\SetFigFont{12}{14.4}{rm}*}}}
\put(170,705){\makebox(0,0)[lb]{\smash{\SetFigFont{12}{14.4}{rm}Q=2}}}
\put(290,705){\makebox(0,0)[lb]{\smash{\SetFigFont{12}{14.4}{rm}Q=3}}}
\put(170,760){\makebox(0,0)[lb]{\smash{\SetFigFont{12}{14.4}{rm}*}}}
\put(170,785){\makebox(0,0)[lb]{\smash{\SetFigFont{12}{14.4}{rm}*}}}
\put(170,805){\makebox(0,0)[lb]{\smash{\SetFigFont{12}{14.4}{rm}*}}}
\put(170,740){\makebox(0,0)[lb]{\smash{\SetFigFont{12}{14.4}{rm}*}}}
\put(310,785){\makebox(0,0)[lb]{\smash{\SetFigFont{12}{14.4}{rm}*}}}
\put(310,760){\makebox(0,0)[lb]{\smash{\SetFigFont{12}{14.4}{rm}*}}}
\put(310,805){\makebox(0,0)[lb]{\smash{\SetFigFont{12}{14.4}{rm}*}}}
\end{picture}
$$
At $\gamma =\pi /7$ the configuration related to $Q=3$ becomes
infinite
$$
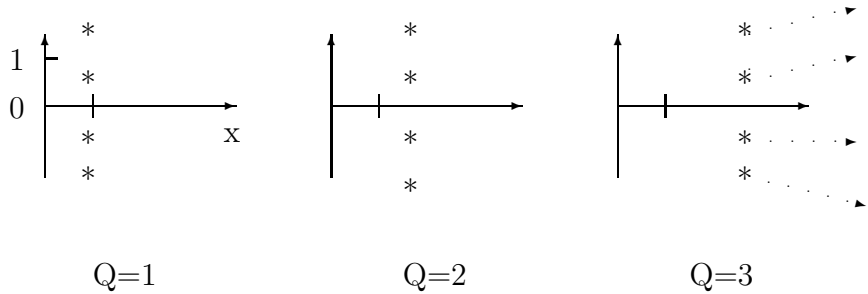
\setlength{\unitlength}{0.012500in}%
\begingroup\makeatletter\ifx\SetFigFont\undefined
% extract first six characters in \fmtname
\def\x#1#2#3#4#5#6#7\relax{\def\x{#1#2#3#4#5#6}}%
\expandafter\x\fmtname xxxxxx\relax \def\y{splain}%
\ifx\x\y   % LaTeX or SliTeX?
\gdef\SetFigFont#1#2#3{%
  \ifnum #1<17\tiny\else \ifnum #1<20\small\else
  \ifnum #1<24\normalsize\else \ifnum #1<29\large\else
  \ifnum #1<34\Large\else \ifnum #1<41\LARGE\else
     \huge\fi\fi\fi\fi\fi\fi
  \csname #3\endcsname}%
\else
\gdef\SetFigFont#1#2#3{\begingroup
  \count@#1\relax \ifnum 25<\count@\count@25\fi
  \def\x{\endgroup\@setsize\SetFigFont{#2pt}}%
  \expandafter\x
    \csname \romannumeral\the\count@ pt\expandafter\endcsname
    \csname @\romannumeral\the\count@ pt\endcsname
  \csname #3\endcsname}%
\fi
\fi\endgroup
\begin{picture}(360,115)(5,705)
\thinlines
\put( 20,750){\vector( 0, 1){ 60}}
\put( 20,780){\vector( 1, 0){ 80}}
\put( 20,800){\line( 1, 0){  5}}
\put( 40,785){\line( 0,-1){ 10}}
\put(140,750){\vector( 0, 1){ 60}}
\put(140,780){\vector( 1, 0){ 80}}
\put(160,785){\line( 0,-1){ 10}}
\put(260,750){\vector( 0, 1){ 60}}
\put(260,780){\vector( 1, 0){ 80}}
\put(280,785){\line( 0,-1){ 10}}
\put(364,738){\vector( 4,-1){0}}
\multiput(315,750)(8.23530,-2.05883){6}{\makebox(0.1111,0.7778){\SetFigFont{5}{6}{rm}.}}
\put(360,765){\vector( 1, 0){0}}
\multiput(315,765)(9.00000,0.00000){5}{\makebox(0.1111,0.7778){\SetFigFont{5}{6}{rm}.}}
\put(315,795){\makebox(0.1111,0.7778){\SetFigFont{5}{6}{rm}.}}
\put(360,801){\vector( 4, 1){0}}
\multiput(315,790)(8.94118,2.23529){5}{\makebox(0.1111,0.7778){\SetFigFont{5}{6}{rm}.}}
\put(315,810){\makebox(0.1111,0.7778){\SetFigFont{5}{6}{rm}.}}
\put(360,821){\vector( 4, 1){0}}
\multiput(315,810)(8.94118,2.23529){5}{\makebox(0.1111,0.7778){\SetFigFont{5}{6}{rm}.}}
\put( 20,780){\line( 1, 0){  5}}
\put( 40,705){\makebox(0,0)[lb]{\smash{\SetFigFont{12}{14.4}{rm}Q=1}}}
\put( 35,745){\makebox(0,0)[lb]{\smash{\SetFigFont{12}{14.4}{rm}*}}}
\put( 95,765){\makebox(0,0)[lb]{\smash{\SetFigFont{12}{14.4}{rm}x}}}
\put(  5,775){\makebox(0,0)[lb]{\smash{\SetFigFont{12}{14.4}{rm}0}}}
\put(  5,795){\makebox(0,0)[lb]{\smash{\SetFigFont{12}{14.4}{rm}1}}}
\put( 35,760){\makebox(0,0)[lb]{\smash{\SetFigFont{12}{14.4}{rm}*}}}
\put( 35,785){\makebox(0,0)[lb]{\smash{\SetFigFont{12}{14.4}{rm}*}}}
\put( 35,805){\makebox(0,0)[lb]{\smash{\SetFigFont{12}{14.4}{rm}*}}}
\put(170,705){\makebox(0,0)[lb]{\smash{\SetFigFont{12}{14.4}{rm}Q=2}}}
\put(290,705){\makebox(0,0)[lb]{\smash{\SetFigFont{12}{14.4}{rm}Q=3}}}
\put(170,760){\makebox(0,0)[lb]{\smash{\SetFigFont{12}{14.4}{rm}*}}}
\put(170,785){\makebox(0,0)[lb]{\smash{\SetFigFont{12}{14.4}{rm}*}}}
\put(170,805){\makebox(0,0)[lb]{\smash{\SetFigFont{12}{14.4}{rm}*}}}
\put(170,740){\makebox(0,0)[lb]{\smash{\SetFigFont{12}{14.4}{rm}*}}}
\put(310,785){\makebox(0,0)[lb]{\smash{\SetFigFont{12}{14.4}{rm}*}}}
\put(310,760){\makebox(0,0)[lb]{\smash{\SetFigFont{12}{14.4}{rm}*}}}
\put(310,805){\makebox(0,0)[lb]{\smash{\SetFigFont{12}{14.4}{rm}*}}}
\put(310,745){\makebox(0,0)[lb]{\smash{\SetFigFont{12}{14.4}{rm}*}}}
\end{picture}
$$
and the allowed integers are reduced to $Q=1,2$. If the anisotropy is
increased to a critical value between $\pi /7>\gamma >\pi /6$,
the imaginary part of one complex pair becomes smaller and vanishes
at the critical value.
This complex pair is replaced by a pair of two real roots.
$$
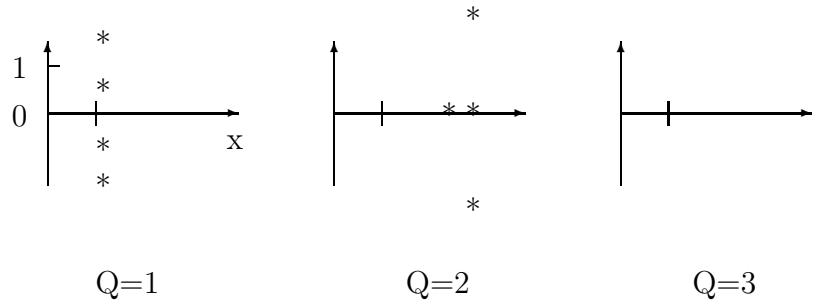
\setlength{\unitlength}{0.012500in}%
\begingroup\makeatletter\ifx\SetFigFont\undefined
% extract first six characters in \fmtname
\def\x#1#2#3#4#5#6#7\relax{\def\x{#1#2#3#4#5#6}}%
\expandafter\x\fmtname xxxxxx\relax \def\y{splain}%
\ifx\x\y   % LaTeX or SliTeX?
\gdef\SetFigFont#1#2#3{%
  \ifnum #1<17\tiny\else \ifnum #1<20\small\else
  \ifnum #1<24\normalsize\else \ifnum #1<29\large\else
  \ifnum #1<34\Large\else \ifnum #1<41\LARGE\else
     \huge\fi\fi\fi\fi\fi\fi
  \csname #3\endcsname}%
\else
\gdef\SetFigFont#1#2#3{\begingroup
  \count@#1\relax \ifnum 25<\count@\count@25\fi
  \def\x{\endgroup\@setsize\SetFigFont{#2pt}}%
  \expandafter\x
    \csname \romannumeral\the\count@ pt\expandafter\endcsname
    \csname @\romannumeral\the\count@ pt\endcsname
  \csname #3\endcsname}%
\fi
\fi\endgroup
\begin{picture}(335,125)(5,705)
\thinlines
\put( 20,750){\vector( 0, 1){ 60}}
\put( 20,780){\vector( 1, 0){ 80}}
\put( 20,800){\line( 1, 0){  5}}
\put( 40,785){\line( 0,-1){ 10}}
\put(140,750){\vector( 0, 1){ 60}}
\put(160,785){\line( 0,-1){ 10}}
\put(260,750){\vector( 0, 1){ 60}}
\put(260,780){\vector( 1, 0){ 80}}
\put(280,785){\line( 0,-1){ 10}}
\put(140,780){\vector( 1, 0){ 80}}
\put( 40,705){\makebox(0,0)[lb]{\smash{\SetFigFont{12}{14.4}{rm}Q=1}}}
\put( 20,780){\line( 1, 0){  5}}
\put( 95,765){\makebox(0,0)[lb]{\smash{\SetFigFont{12}{14.4}{rm}x}}}
\put( 40,805){\makebox(0,0)[lb]{\smash{\SetFigFont{12}{14.4}{rm}*}}}
\put(  5,775){\makebox(0,0)[lb]{\smash{\SetFigFont{12}{14.4}{rm}0}}}
\put(  5,795){\makebox(0,0)[lb]{\smash{\SetFigFont{12}{14.4}{rm}1}}}
\put(170,705){\makebox(0,0)[lb]{\smash{\SetFigFont{12}{14.4}{rm}Q=2}}}
\put(290,705){\makebox(0,0)[lb]{\smash{\SetFigFont{12}{14.4}{rm}Q=3}}}
\put(195,815){\makebox(0,0)[lb]{\smash{\SetFigFont{12}{14.4}{rm}*}}}
\put(195,735){\makebox(0,0)[lb]{\smash{\SetFigFont{12}{14.4}{rm}*}}}
\put(185,775){\makebox(0,0)[lb]{\smash{\SetFigFont{12}{14.4}{rm}*}}}
\put(195,775){\makebox(0,0)[lb]{\smash{\SetFigFont{12}{14.4}{rm}*}}}
\put( 40,760){\makebox(0,0)[lb]{\smash{\SetFigFont{12}{14.4}{rm}*}}}
\put( 40,745){\makebox(0,0)[lb]{\smash{\SetFigFont{12}{14.4}{rm}*}}}
\put( 40,785){\makebox(0,0)[lb]{\smash{\SetFigFont{12}{14.4}{rm}*}}}
\end{picture}
$$
Thus, the 4-string degenerates. The whole configuration can be viewed
as a set of one finite 1-string and one 3-string which tends to infinity at
$\gamma\to\pi /6$.
$$
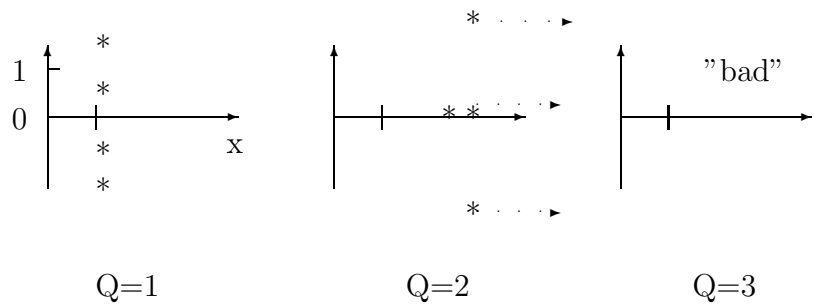
\setlength{\unitlength}{0.012500in}%
\begingroup\makeatletter\ifx\SetFigFont\undefined
% extract first six characters in \fmtname
\def\x#1#2#3#4#5#6#7\relax{\def\x{#1#2#3#4#5#6}}%
\expandafter\x\fmtname xxxxxx\relax \def\y{splain}%
\ifx\x\y   % LaTeX or SliTeX?
\gdef\SetFigFont#1#2#3{%
  \ifnum #1<17\tiny\else \ifnum #1<20\small\else
  \ifnum #1<24\normalsize\else \ifnum #1<29\large\else
  \ifnum #1<34\Large\else \ifnum #1<41\LARGE\else
     \huge\fi\fi\fi\fi\fi\fi
  \csname #3\endcsname}%
\else
\gdef\SetFigFont#1#2#3{\begingroup
  \count@#1\relax \ifnum 25<\count@\count@25\fi
  \def\x{\endgroup\@setsize\SetFigFont{#2pt}}%
  \expandafter\x
    \csname \romannumeral\the\count@ pt\expandafter\endcsname
    \csname @\romannumeral\the\count@ pt\endcsname
  \csname #3\endcsname}%
\fi
\fi\endgroup
\begin{picture}(335,125)(5,705)
\thinlines
\put( 20,750){\vector( 0, 1){ 60}}
\put( 20,780){\vector( 1, 0){ 80}}
\put( 20,800){\line( 1, 0){  5}}
\put( 40,785){\line( 0,-1){ 10}}
\put(140,750){\vector( 0, 1){ 60}}
\put(160,785){\line( 0,-1){ 10}}
\put(260,750){\vector( 0, 1){ 60}}
\put(260,780){\vector( 1, 0){ 80}}
\put(280,785){\line( 0,-1){ 10}}
\put(140,780){\vector( 1, 0){ 80}}
\put(235,740){\vector( 1, 0){0}}
\multiput(200,740)(8.75000,0.00000){4}{\makebox(0.1111,0.7778){\SetFigFont{5}{6}{rm}.}}
\put(240,820){\vector( 1, 0){0}}
\multiput(200,820)(10.00000,0.00000){4}{\makebox(0.1111,0.7778){\SetFigFont{5}{6}{rm}.}}
\put(235,785){\vector( 1, 0){0}}
\multiput(200,785)(8.75000,0.00000){4}{\makebox(0.1111,0.7778){\SetFigFont{5}{6}{rm}.}}
\put( 20,780){\line( 1, 0){  5}}
\put( 40,705){\makebox(0,0)[lb]{\smash{\SetFigFont{12}{14.4}{rm}Q=1}}}
\put(295,795){\makebox(0,0)[lb]{\smash{\SetFigFont{12}{14.4}{rm}"bad"}}}
\put( 95,765){\makebox(0,0)[lb]{\smash{\SetFigFont{12}{14.4}{rm}x}}}
\put(  5,775){\makebox(0,0)[lb]{\smash{\SetFigFont{12}{14.4}{rm}0}}}
\put(  5,795){\makebox(0,0)[lb]{\smash{\SetFigFont{12}{14.4}{rm}1}}}
\put(170,705){\makebox(0,0)[lb]{\smash{\SetFigFont{12}{14.4}{rm}Q=2}}}
\put(290,705){\makebox(0,0)[lb]{\smash{\SetFigFont{12}{14.4}{rm}Q=3}}}
\put(195,815){\makebox(0,0)[lb]{\smash{\SetFigFont{12}{14.4}{rm}*}}}
\put(195,735){\makebox(0,0)[lb]{\smash{\SetFigFont{12}{14.4}{rm}*}}}
\put(185,775){\makebox(0,0)[lb]{\smash{\SetFigFont{12}{14.4}{rm}*}}}
\put(195,775){\makebox(0,0)[lb]{\smash{\SetFigFont{12}{14.4}{rm}*}}}
\put( 40,760){\makebox(0,0)[lb]{\smash{\SetFigFont{12}{14.4}{rm}*}}}
\put( 40,745){\makebox(0,0)[lb]{\smash{\SetFigFont{12}{14.4}{rm}*}}}
\put( 40,785){\makebox(0,0)[lb]{\smash{\SetFigFont{12}{14.4}{rm}*}}}
\put( 40,805){\makebox(0,0)[lb]{\smash{\SetFigFont{12}{14.4}{rm}*}}}
\end{picture}
$$
In this case only $Q=1$ is permitted. However, in the limit
$\gamma\to\pi /5$ this set degenerates into two (2-string like) complex
pairs. One 2-string is divergent.
$$
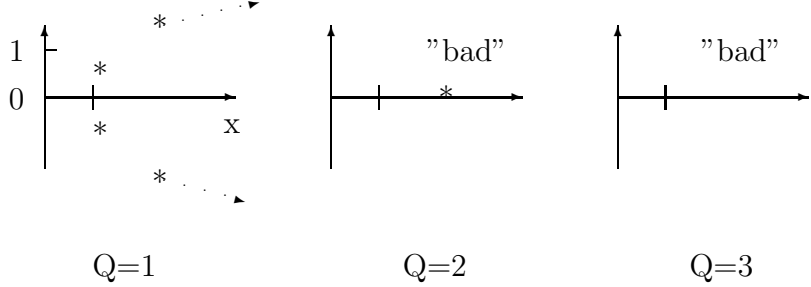
\setlength{\unitlength}{0.012500in}%
\begingroup\makeatletter\ifx\SetFigFont\undefined
% extract first six characters in \fmtname
\def\x#1#2#3#4#5#6#7\relax{\def\x{#1#2#3#4#5#6}}%
\expandafter\x\fmtname xxxxxx\relax \def\y{splain}%
\ifx\x\y   % LaTeX or SliTeX?
\gdef\SetFigFont#1#2#3{%
  \ifnum #1<17\tiny\else \ifnum #1<20\small\else
  \ifnum #1<24\normalsize\else \ifnum #1<29\large\else
  \ifnum #1<34\Large\else \ifnum #1<41\LARGE\else
     \huge\fi\fi\fi\fi\fi\fi
  \csname #3\endcsname}%
\else
\gdef\SetFigFont#1#2#3{\begingroup
  \count@#1\relax \ifnum 25<\count@\count@25\fi
  \def\x{\endgroup\@setsize\SetFigFont{#2pt}}%
  \expandafter\x
    \csname \romannumeral\the\count@ pt\expandafter\endcsname
    \csname @\romannumeral\the\count@ pt\endcsname
  \csname #3\endcsname}%
\fi
\fi\endgroup
\begin{picture}(335,115)(5,705)
\thinlines
\put( 20,750){\vector( 0, 1){ 60}}
\put( 20,780){\vector( 1, 0){ 80}}
\put( 20,800){\line( 1, 0){  5}}
\put( 40,785){\line( 0,-1){ 10}}
\put(140,750){\vector( 0, 1){ 60}}
\put(160,785){\line( 0,-1){ 10}}
\put(260,750){\vector( 0, 1){ 60}}
\put(260,780){\vector( 1, 0){ 80}}
\put(280,785){\line( 0,-1){ 10}}
\put(140,780){\vector( 1, 0){ 80}}
\put(104,736){\vector( 4,-1){0}}
\multiput(
70,745)(8.52940,-2.13235){4}{\makebox(0.1111,0.7778){\SetFigFont{5}{6}{rm}.}}
\put(110,820){\vector( 4, 1){0}}
\multiput(
70,810)(8.00000,2.00000){5}{\makebox(0.1111,0.7778){\SetFigFont{5}{6}{rm}.}}
\put( 20,780){\line( 1, 0){  5}}
\put( 40,705){\makebox(0,0)[lb]{\smash{\SetFigFont{12}{14.4}{rm}Q=1}}}
\put( 65,740){\makebox(0,0)[lb]{\smash{\SetFigFont{12}{14.4}{rm}*}}}
\put( 95,765){\makebox(0,0)[lb]{\smash{\SetFigFont{12}{14.4}{rm}x}}}
\put(  5,775){\makebox(0,0)[lb]{\smash{\SetFigFont{12}{14.4}{rm}0}}}
\put(  5,795){\makebox(0,0)[lb]{\smash{\SetFigFont{12}{14.4}{rm}1}}}
\put(170,705){\makebox(0,0)[lb]{\smash{\SetFigFont{12}{14.4}{rm}Q=2}}}
\put(290,705){\makebox(0,0)[lb]{\smash{\SetFigFont{12}{14.4}{rm}Q=3}}}
\put(185,775){\makebox(0,0)[lb]{\smash{\SetFigFont{12}{14.4}{rm}*}}}
\put( 40,760){\makebox(0,0)[lb]{\smash{\SetFigFont{12}{14.4}{rm}*}}}
\put( 40,785){\makebox(0,0)[lb]{\smash{\SetFigFont{12}{14.4}{rm}*}}}
\put(295,795){\makebox(0,0)[lb]{\smash{\SetFigFont{12}{14.4}{rm}"bad"}}}
\put(180,795){\makebox(0,0)[lb]{\smash{\SetFigFont{12}{14.4}{rm}"bad"}}}
\put( 65,805){\makebox(0,0)[lb]{\smash{\SetFigFont{12}{14.4}{rm}*}}}
\end{picture}
$$
Therefore, no integer is allowed for a 4-string
($\pi /\gamma =r\leq 5$).

These examples and similar ones lead us to the conjectures
about admissible configurations of finite strings of positive parity
stated in section \ref{sec:bae-string}.

\section*{Appendix C: Proof of the Lemma in Section \ref{sec:bae-string}}

We define the function
\begin{equation}
\tilde Z(N,l,\mu,r)=\sum_{k=-\infty}^\infty (f_{k,1}-f_{k,0})
\label{z-tilde}
\end{equation}
and     show   that       it  solves    the      recurrence   relation
(\ref{recurrence-relation})   and  fulfills   the  initial   condition
(\ref{initial-condition}).  Only  a finite number of  terms contribute
because the binomial coefficients ${m\choose n}$ are zero for $m<0$ or
$n<0$.

The right hand side of the recurrence relation reads for $f_{k,d}$
\begin{eqnarray}
F_{k,d}(N,l,\mu,r)&=&
\sum_{\nu_0 =0}^{\mu -1}{N-2\mu +\nu_0-G_0(r)\choose \nu_0}
             f_{k,d}(N-2\mu,l-\mu,\mu-\nu_0,r-1) \\
&=&\sum_{\nu_0 =0}^{\mu -1}{N-2\mu +\nu_0-G_0(r)\choose \nu_0}
               {N-l-k(r-3)-G_0(r)+d-\mu\choose
                N-l+1-k(r-2)-2\mu -G_0(r)+\nu_0}\nonumber\\
&&    \times    {l+k(r-3)-d-\mu\choose
                l+k(r-2)-2\mu+\nu_0}
\end{eqnarray}
Using the sum rule of Binomial coefficients
\begin{equation}
{a+b\choose n}=\sum_{k=0}^n {a\choose k}{b\choose n-k}
\label{binomial-sum-rule}
\end{equation}
one has (with $g=G_0(r)$ and $\nu =\nu_0$)
\begin{equation}
  {N-2\mu-g+\nu\choose\nu}=\sum_{\omega =0}^\nu
  {N-2\mu+\nu-l-g+1-k(r-2)\choose \nu -\omega}
  {l-1+k(r-2)\choose\omega}
\end{equation}
which is inserted in $F_{k,d}$. With the identity
\begin{eqnarray}
&&  {N-2\mu+\nu-l-g+1-k(r-2)\choose \nu -\omega}
    {N-l-g+d-k(r-3)-\mu\choose N-l+1-g-k(r-2)-2\mu+\nu} \nonumber\\
&=&  {N-l-g+d-k(r-3)-\mu\choose d+k+\mu-1-\omega}
    {d+k+\mu-1-\omega\choose \nu-\omega}
\end{eqnarray}
we have
\begin{eqnarray}
F_{k,d} &=& \sum_{\omega =0}
  {l-1+k(r-2) \choose \omega}
  {N-l+d-g-k(r-3)-\mu \choose d+k+\mu-1-\omega}\times \nonumber\\
&&\sum_{\nu =\omega}
  {d+k+\mu-1-\omega\choose\nu-\omega}
  {l-d+k(r-3)-\mu\choose l+k(r-2)-2\mu+\nu}.
\end{eqnarray}
With the help of the sum rule (\ref{binomial-sum-rule}) this equation
reads
\begin{eqnarray}
F_k &=& \sum_{\omega =0}
  {l-1+k(r-2) \choose \omega}
  {N-l+d-g-k(r-3)-\mu \choose d+k+\mu-1-\omega}\times\nonumber\\
&&  {l-1+k(r-2)-\omega\choose l-1+k(r-1)+d-\mu} \\
&=& \sum_{\omega =0}
  {l-1+k(r-2) \choose \mu-d-k}
  {\mu-d-k\choose\omega}
  {N-l+d-g-k(r-3)-\mu \choose d+k+\mu-1-\omega}
\end{eqnarray}
with the final result
\begin{equation}
F_{k,d}={N-l-k(r-2)-g\choose N-l+1-k(r-1)-g-\mu-d}
    {l-1+k(r-2)\choose l+k(r-1)-\mu-1+d}.
\end{equation}
In  the   ``good''    sector   $2j+1<r-1$,   where   by   (\ref{qmax})
$G_0(r)=G_0(r+1)=g=0$ it is    easy to show the  recurrence   relation
(\ref{recurrence-relation})  for   term in (\ref{recurrence-solution})
separately
\begin{equation}
F_{k,1}-F_{k,0}=f_{k,1}-f_{k,0}.
\end{equation}
Thus the  function $\tilde Z(N,l,\mu,r)$ (\ref{z-tilde})  fulfills the
recurrence   relation.   But in  order to   use the initial conditions
(\ref{initial-condition}) and  (\ref{initial-condition-r=2})  we  also
have to  consider the   sector  $2j+1\geq r-1$ where    $G_0(r)>0$  and
$G_0(r)=G_0(r+1)+1$ which yields additional terms
\begin{equation}
F_{k,1}-F_{k,0} = f_{k,1}-f_{k,0}+ K_{k,1}-K_{k,0}
\end{equation}
with
\begin{equation}
K_{k,d} =
    {N-l-k(r-2)-G_0(r+1)-d\choose N-l+1-k(r-1)-G_0(r+1)-\mu}
    {l+k(r-2)+d-1\choose l+k(r-1)-\mu}.
\end{equation}
The   function  $\tilde Z$   (\ref{z-tilde}) fulfills  the  recurrence
relation also for $2j+1\geq r-1$ due to the following identity
\begin{equation}
  \sum_k \left( K_{k,1}-K_{k,0}\right) =0.
\label{identity-k-sum}
\end{equation}
We have no analytic  proof of this  identity  but we have verified  it
numerically for all possible cases for $N<60$. It is easy to show that
$\tilde Z$ assume the initial conditions.
Therefore, the function $\tilde Z(N,l,\mu,r)=Z(N,l,\mu,r)$
is equal to the partial number of states defined in
eq.~(\ref{z-string-sum}).

\end{appendix}

\renewcommand{\baselinestretch}{1}\small\normalsize
\section*{\refname}
\begin{list}{}{\leftmargin0cm}

\item Alcaraz A G, Barber M N, Batchelor M T, Baxter R J and Quispel G
  R W 1987 {\em J. Phys. A: Math. Gen.} {\bf 20} 6397

\item Andrews  G E, Baxter  R J and Forrester  P J  1984 {\em J. Stat.
    Phys} {\bf 35} 193

\item Baxter R   J  1982 {\em Exactly   Solved  Models in  Statistical
    Mechanics} (New York: Academic)

\item Bethe H A 1931 {\em Z. Phys} {\bf 71} 205

\item Bo-Yu  Hou, Kang-Jie Shi, Zhong-Xia Yang  and  Rui-Hong Yue 1991
  {\em J. Phys. A: Math. Gen.} {\bf 24} 3825

%\item
%Cardy J L 1984  {\em J.Phys.A: Math.Gen.} {\bf 17} L385
%
%\item
%Cardy J L 1987  {\em Phase Transitions and  Critical  Phenomena}  Vol  11
%                  Ed. C Domb and J L Lebowitz (London: Academic)
%

\item Cherednik I 1984 {\em Theor. Mat. Fiz.} {\bf 61} 35

\item Destri C, de Vega H J 1992 {\em Nucl. Phys.} {\bf B385} 361

\item Drinfeld V G 1986 {\em Proc. Int. Cong. Math.} Berkeley

\item E\ss  ler  H L,  Korepin V E   and Schoutens K 1992a {\em  Nucl.
    Phys.} {\bf B384} 431

\item E\ss ler H L, Korepin V E and Schoutens K 1992b {\em J. Phys. A:
    Math. Gen.} {\bf 25} 4115

%\item
%Faddeev L D and Takhtajan 1981 {\em Zap. Nauch. Semin.}
%    LOMI Vol 109 p. 134

\item Foerster A and Karowski M 1992 {\em Phys. Rev. B} {\bf 46} 9234

\item Foerster A and Karowski M 1993 {\em Nucl. Phys.} {\bf B396} 611

\item Foerster A and Karowski M 1993 {\em Nucl. Phys.} {\bf B408} 512

\item Friedan D, Qiu Z and Shenker S 1984  {\em Phys. Rev. Lett.} Vol.
  52 {\bf 18} 1575

%\item
%Gaudin M 1983 {\em La Fonction d'onde de Bethe} Serie Scientifique du
%CEA (Paris: Masson)

\item Hamer C J, Quispel G R W and Batchelor M T 1987 {\em J. Phys. A:
    Math. Gen.} {\bf 20} 5677

\item Jimbo M 1985 {\em Lett. Math. Phys.} {\bf 10} 63

\item J\"uttner G and  D\"orfel B-D 1993 {\em  J. Phys. A: Math. Gen.}
  {\bf 26} 3105

\item Karowski M 1988 {\em Nucl. Phys.} {\bf B300} 473

%\item
%Karowski M and Schrader R 1993 {\em Commun. Math. Phys.}{\bf 151} 355

\item Karowski  M and  Zapletal  A 1993 {\em Quantum   Group Invariant
    Integrable n-State    Vertex   Models   with  Periodic    Boundary
    Conditions}  preprint   hep-th/9312008, published  in  {\em  Nucl.
    Phys.} {\bf B419}[FS] (1994) 567

\item Karowski M and Zapletal  A 1994 {\em Highest weight $U_q[sl(n)]$
    modules and   invariant integrable n-state   models with  periodic
    boundary conditions} preprint hep-th/9406116

\item Kirillov A N and  Liskova N A 1994  {\em Completeness of Bethe's
    States for Generalized XXZ Model} preprint hep-th/9403107

\item Kirillov A N and Reshetikhin N Y 1989 {\em Representation of the
    algebra  $U_q(sl(2))$, q-orthogonal polynomials  and invariants of
    links} In: Kohno  T (ed.)  {\em New  developments in the theory of
    knots} Advanced series in Math. Physics, Vol. 11 (Singapore: World
  Scientific)

\item Lusztig E 1989 {\em Contemp. Math.} {\bf 82} 59

\item Martin P P and Rittenberg V 1992 {\em Int.   J. Phys.} {\bf A7},
  Suppl. 1B, 707

\item Mezincescu L and Nepomechie R I 1991 {\em Mod. Phys. Lett.} {\bf
    A6} 2497, {\em Int. J. Phys.} {\bf A6} (1991) 5231

%\item
%Meljanac S, Milekovic M and Pallua S 1991 {\em J. Phys. A: Math. Gen.}
%{\bf 24} 581

\item Pasquier N and Saleur H 1990 {\em Nucl. Phys.} {\bf B330} 523

\item Reshetikin  N  and Smirnov  F 1989  {\em  Hidden  Quantum  Group
    Symmetry and Integrable Perturbations of Conformal Field Theories}
  preprint HUTMP   89/B246, published in  {\em Comm.  Math. Phys.} 131
  (1990) 157

\item Reshetikin N and Turaev V G 1991 {\em Invent. math.} 103, 547

\item Rosso M 1988 {\em Commun. Math. Phys.} {\bf 117} 581

\item Sklyanin E K 1988 {\em J. Phys. A: Math. Gen.} {\bf 21} 2375

\item Takahashi M 1971 {\em Prog.Theor.Phys.} {\bf 46} 401

\item Takahashi M 1972 {\em Prog.Theor.Phys.} {\bf 48} 2187

\item Vladimirov A A 1984 {\em Non-String Two-Magnon Configurations in
    the  Isotropic Heisenberg  Magnet}   Joint Institute  for  Nuclear
  Research Dubna preprint P17-84-409

\end{list}
\end{document}